\documentclass[aps,prl,reprint,longbibliography]{revtex4-1}

\usepackage[T1]{fontenc}
\usepackage{xr}
\usepackage{amsmath}% Include figure files 
\usepackage{multirow}
\usepackage{graphicx}% Include figure files
\usepackage{color}% Include figure files
\usepackage{dcolumn}% Align table columns on decimal point
\usepackage{bm}% bold math
\graphicspath{{IMAGES/}}

\begin{document}

\preprint{APS/123-QED}

%\author{~~~~~~}
%\affiliation{Department of Physics, Sapienza Universit\`a di Roma, Piazzale Aldo Moro, 2, 00185 Rome, Italy}

%`` <-- use this symbol for left quotes
\author{Lorenzo Rovigatti} 
\affiliation{Department of Physics, {\textit Sapienza} Universit\`a di Roma, Piazzale A. Moro 2, IT-00185 Roma, Italy}
\affiliation{CNR-ISC Uos Sapienza, Piazzale A. Moro 2, IT-00185 Roma, Italy}

\author{Francesco Sciortino}
\affiliation{Department of Physics, {\textit Sapienza} Universit\`a di Roma, Piazzale A. Moro 2, IT-00185 Roma, Italy}
%\SectionNumbersOn

%%%%%%%%%%%%%%%%%%%%%%%%%%%%%%%%%%%%%%%%%%%
%%%%%%%%%%%%%%%%%%%%%%%%%%%%%%%%%%%%%%%%%%%

\date{\today}

\begin{abstract}
Single-chain nanoparticles (SCNP) 
are a new class of bio and soft-matter polymeric objects in which  a fraction of the monomers are able to form equivalently intra- or inter-polymer  bonds.
 % We numerically show that 
 Here we numerically  show that a fully-entropic gas-liquid phase separation  can take place in SCNP systems.
  Control over the  discontinuous (first-order) change  --- from a phase of independent diluted (fully-bonded) polymers to a phase in which  polymers entropically bind to each other to form a (fully-bonded)  polymer network --- can be achieved by a judicious design of the patterns of reactive monomers along the polymer chain.  Such a sensitivity arises from a delicate balance between the distinct entropic contributions controlling the binding. 
  \end{abstract}

\title{Designing enhanced entropy binding in single-chain nano particles}

\maketitle

{\bf Introduction:}
As elegantly epitomised  in the van der Waals theory~\cite{vdw}, in atomic systems the gas-liquid phase separation phenomenon originates from inter-particle attraction. 
More recently, colloidal systems have provided evidence of purely entropy-driven   
 ``gas-liquid'' phase transitions, as observed  in the presence of  depletion interactions~\cite{asakura1954interaction,tuinier2011colloids}, combinatorial attractions~\cite{zilman2003entropic} and hard-core interactions between particles with specific shapes~\cite{lee2019entropic}.
 
 While depletion interactions have been studied in details in the last 30 years~\cite{tuinier2011colloids,doi:10.1063/5.0049350}, combinatorial attractions have received much less attention.   In their seminal study, Safran  and coworkers~\cite{zilman2003entropic} investigated a system composed by
microemulsion droplets linked by telechelic polymers~\cite{filali2001robust}.  The polymer body is exposed to  the aqueous solvent while the hydrophobic ends are  constrained to reside inside the same or in two distinct oil droplets.   The different ways the polymer ends can be distributed
over the accessible droplets   leads to  droplet condensation, i.e. to the liquid state. DNA-coated colloids provide further examples in which combinatorial entropy can be exploited to drive phase separation. Here particles are grafted with equal quantities of sticky ends and of their complementary sequence~\cite{angioletti2012re,bachmann2016melting}, or grafted with palindromic  sequences~\cite{sciortino2019entropy,sciortino2020combinatorial}. In both cases, complete DNA hybridization can take place inside the same particle or between distinct particles and the balance between these two possibilities is controlled by a combinatorial entropic contribution.

In both telechelic polymers and DNA coated particles 
the dominant inter-particle contributions are strong interactions
 of the  lock and key type such that the system is constantly in its energetic ground  (fully-bonded) state. Particles  can
 satisfy all possible  bonds  both in  the gas phase (colloid poor) via intra-particle bonds as well as  in the liquid phase (colloid rich), where bonds are shared between different particles. 
Being the number of bonds (and hence the energy) the same in both phases, entropy becomes the only  driving force for condensation~\cite{smallenburg2013liquids}.

Functionalized polymers, in which a fraction of the monomers are able to form reversible bonds, have recently entered the radar of the soft-matter~\cite{kumar2001gelation,seo2008polymeric,lyon2015brief,arbe2016structure, pomposo2017single,whitaker2013thermoresponsive,tang2015anomalous,ghosh2020physical,oyarzun2018programming} and biophysics~\cite{statt2020model,ruff2021ligand,lichtinger2021targeted,dignon2018relation,ZUMBRO2019892} communities, and  experimentally  synthesised even inside cells to promote gelation~\cite{nakamura2018intracellular}.  
If the chain flexibility is large enough and the associative monomers can form only single bonds, then at low density bonding takes place essentially within the same polymer, forming soft nano-objects named single-chain nanoparticles (SCNP)~\cite{pomposo2017single}. At larger densities the combinatorial entropy should favour  phase separation. However, in this case the free energy has additional terms that stem from the polymeric nature of the nanoparticles. Indeed, in contrast to colloids, where in addition to bonding the only other contribution is provided by the steric repulsion, in SNCP systems one has to also take into account the conformational entropy contribution associated to the change from an intra-polymer to an inter-polymer bond. The dependence of all these entropic terms on the number and type of attractive sites is complex~\cite{moreno2017effect} and has not been completely mapped out yet. In general, the interplay between the entropic contributions in play is subtle, and the resulting phase behaviour difficult to predict. In the specific case of SCNPs, at high density no hints of a first-order  transition have been  observed in experiments~\cite{whitaker2013thermoresponsive,tang2015anomalous}  and simulations~\cite{formanek2021gel}, consistent with predictions of  mean-field theory~\cite{semenov1998thermoreversible}. By contrast, a continuous cross-over from isolated chains to percolating states have been observed~\cite{whitaker2013thermoresponsive,formanek2021gel,tang2015anomalous}, akin to the gelation without phase separation phenomenon observed in polymer and biopolymer systems~\cite{flory1942constitution,harmon2017intrinsically}.

Here we show that, opposite to what previously found and thought, a fully-entropic gas-liquid phase separation can take place in SCNP systems.  By studying a series of differently functionalized polymers we demonstrate that phase separation in this system takes place 
due to both the attractive combinatorial entropy and the conformational entropy contribution associated to the change from an intra-polymer to an inter-polymer bond.   We show that this last term can be modulated by designing the sequence of reactive monomers, offering the possibility to discontinuously change, preserving all bonds, from a dilute gas of 
independent polymers to a phase in which different polymers bind to each other to form an extended network.

We perform molecular dynamics simulations of Kremer-Grest polymers~\cite{kremer1990dynamics} complemented by attractive monomers that interact through a potential that  enforces the single-bond per reactive monomer condition and enables a bond-swapping mechanism.
The algorithm~\cite{sciortino2017three}, recently applied to a variety of soft-matter systems~\cite{gnan2017silico,rovigatti2018self,ciarella2018dynamics,sorichetti2021effect},  is capable to modify the bonding pattern close to the fully bonded state,  even when the thermal energy $k_BT \equiv 1/\beta$ is much smaller than the bond strength $\epsilon_b$, overcoming kinetic bottlenecks (see section~S1).

\begin{figure}[htbp] %  figure placement: here, top, bottom, or page
   {\centering
      \includegraphics[width=0.4\textwidth,angle=0]{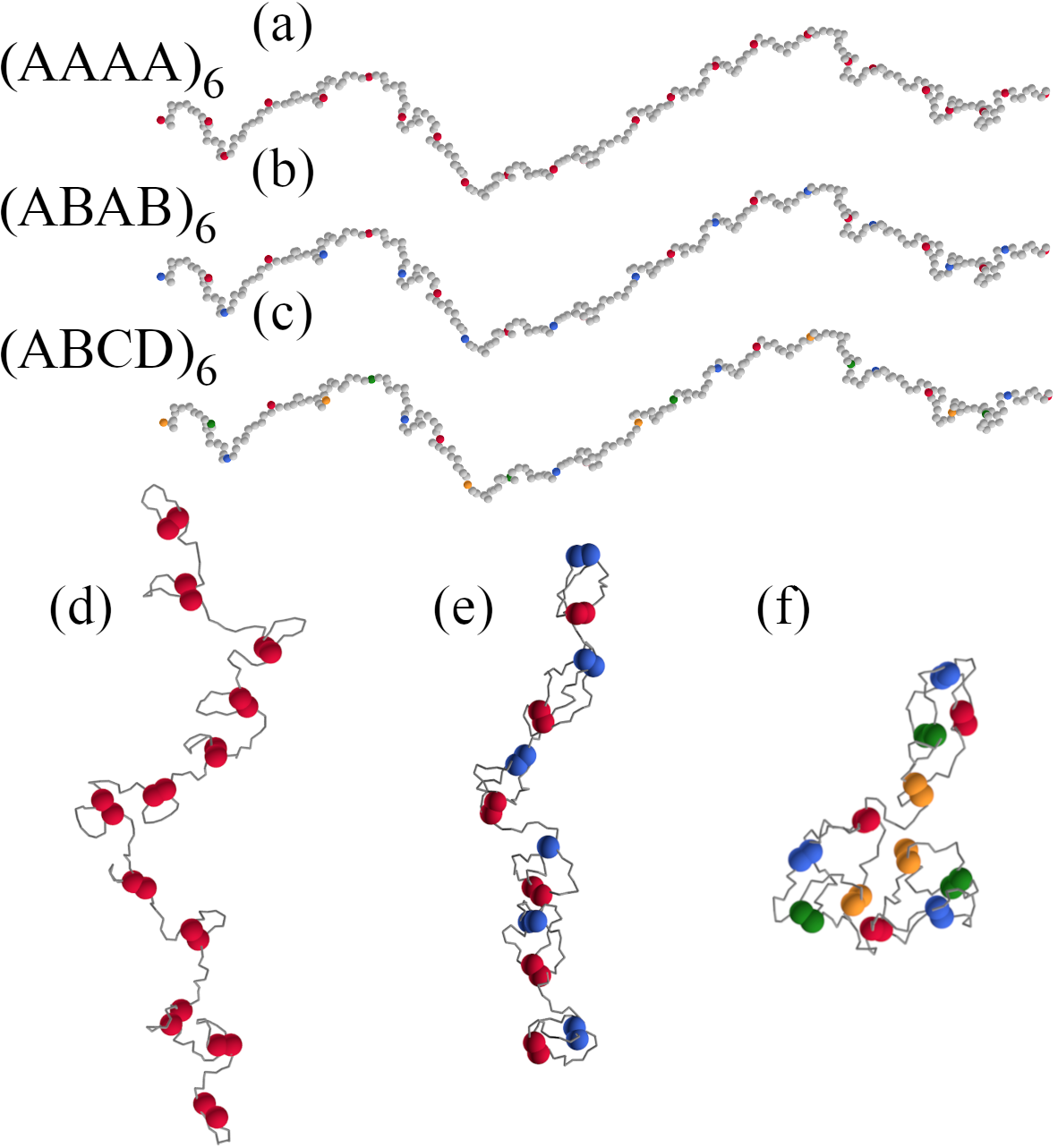} 
             }                                  
   \caption{Cartoons of the investigated polymers highlighting the different conformational changes associated to the formation of a fully bonded configuration in the three models considered. Here inert monomers are coloured in grey, while
    reactive monomers are depicted as  coloured spheres. The three polymers are shown in an open, (a-c) and closed, (d-f), conformation. In (d-f) the inert monomers are not explicitly shown and 
 the reactive monomers are shown with enhanced size for the sake of visibility.   }
   \label{fig:polymers}
\end{figure}

We simulate  polymers  composed by $N_m=254$ monomers, 24 of which are equispaced reactive and 230 inert.
We study a model, named (AAAA)$_6$,  in which all ($A$-type) reactive monomers are identical, a model in which 
$A$ and $B$ reactive monomers alternate, (ABAB)$_6$, and a model with four different ($A$, $B$, $C$, $D$)  alternating reactive monomers, (ABCD)$_6$.   A cartoon of the three studied  polymer models  is shown in Fig.~\ref{fig:polymers}(a-c). 
 Reactive monomers are able to form one and only one strong bond with another same-type monomer. 
We perform two types of calculations. In the first, we simulate two polymers  to compute the effective potential as a function of the relative distance between their centers of mass, while in the second one we perform bulk simulations  of hundred or more polymers with periodic boundary conditions to compute the equation of state and the coexistence between phases.  Further information on the numerical methods are available in Ref.~\cite{sm}.

{\bf The problem:}
Consider a SCNP, a polymer in which $N_R$ of the constituent monomers are reactive. Each reactive monomer 
can form a strong bond with another reactive monomer of the same type
on the same or on a nearby polymer. 
Different from  functionalized colloidal (patchy) particles~\cite{bianchi2006phase,sciortino2017equilibrium}, in which the rigidity of the  particle prevents reactive monomers belonging to the same object	 from bonding with each other, in the polymer case all reactive monomers 
 can  take part in intra-polymer bonds.  Being $\epsilon_b \gg k_BT$, each polymer can assume a fully bonded (ground state) configuration  in which it is disconnected from all other polymers (Fig.~\ref{fig:polymers}(d-f)).
This raises the question whether this ``independent-polymer''  state is the highest entropy state for a system of such polymers or if  
swapping  intra-polymer for inter-polymer bonds  can increase the system entropy even further. Even more important is the question about whether the increase in entropy, if present, is strong enough to induce condensation of a dense ``liquid'' phase starting from a dilute polymer solution.

{\bf Effective Potential (expectations):}    We begin investigating the effective potential $\beta V_{\rm eff}(R)$ between two polymers as a function of the  relative center-to-center distance $R$, when  $\epsilon_b  \gg k_BT$.
In this limit, the possible available configurations of  two polymers are restricted to
the ones in which all possible bonds are formed (either intra- or inter-polymer bonds).  Being the total number of bonds always fixed, energy does not play any role in the interaction, leaving entropy as the only driving force.
Three different entropic contributions determine $\beta V_{\rm eff}(R)$.
The first contribution includes the cost of bringing two fully-bonded polymers at relative distance $R$ when only intra-polymer bonds are present. This is the standard polymer-polymer entropic repulsion~\cite{grosberg1982polymeric,bolhuis2001accurate,likos2001effective}. The second contribution is a combinatorial term, which account for the entropy gain of swapping intra- with inter-polymer bonds.  
The number of configurations in which  both intra and inter bonds are allowed is larger 
compared to the case in which only intra-polymer bonds are present (Section~S4), resulting in an attractive contribution~\cite{semenov1998thermoreversible,sciortino2020combinatorial}. 

   The third and last contribution ($S_{\rm conf}$) is linked to the conformational change of the
polymer on going from the all intra-bond conformation to the mixed intra-/inter-
bonds case. This entropic change accounts for the different number of 
configurations available to the inert monomers when 
the bonding pattern changes. $S_{\rm conf}$ is sensitive to
the relative distances between identical reactive monomers along the polymers and hence 
it can be tuned to control the strength of $\beta V_{\rm eff}$ by  changing
the types of the reactive monomers along the chain. 
Interestingly, as discussed in Section~S2 simply changing the number of inert monomers  while leaving the type and number of reactive monomers invariant does not modify the effective potential if distances are rescaled by the gyration radius.

Fig.~\ref{fig:polymers}(a-c) shows the three SCNP discussed here in an open configuration, while 
  Fig.~\ref{fig:polymers}-(d-e-f) shows the same models  in a fully bonded configuration, to highlight the importance of the conformational entropic contribution. 
   The figure vividly shows that  on increasing the number of  distinct reactive types, the fully bonded polymer becomes more compact.  Assuming that   bonds between nearest reactive monomers are the ones preferentially formed~\cite{semenov1998thermoreversible} (an hypothesis  supported by  simulations),  to a first approximation
each polymer in the bonded state can be visualised as an
independent ``unit'' of paired reactive monomers, where the number of independent units is controlled by the number of reactive monomers of the same type. 
The change in conformational entropy $\Delta S_{\rm o\rightarrow fb}$ of single chains going from  an open unbonded state 
(identical for all polymers) to a fully bonded state (different for each of the three polymers considered) can be calculated \textit{via} Hamiltonian integration (Sec.~S1-F). 
The results, reported in Table~\ref{tbl:entropy}, 
 confirm the progressive entropic cost of constraining the  polymer into a configuration in which all bonds are formed on going from  (AAAA)$_6$ to (ABAB)$_6$. 
 Differences of about 1 $k_BT$ per reactive monomer characterize the  (ABAB)$_6$ and the  (ABCD)$_6$  polymers as compared to the (AAAA)$_6$  polymer, a significant  configurational entropic cost required to satisfy all bond constraints, which can be 
 partially regained when intra bonds are swapped with inter-polymer bonds.

\begin{table}[htp]
\begin{center}
\begin{tabular}{ |c|c|c|c| } 
 \hline
       Polymer    & $\Delta S_{\rm o\rightarrow fb}/k_B$  &  $\Delta S_{\rm o\rightarrow fb}/k_B$ & $R_g^2/\sigma^2$  \\
              type    & ~  & per reactive site & ~ \\
\hline
       \hline
       (AAAA)$_6$  &  -93.0  &  -3.87 & 80\\
       \hline
       (ABAB)$_6$ &  -110.6  & -4.60 & 55 \\
        \hline
        (ABCD)$_6$ &   -119.3 & -4.97  & 50 \\
\hline
\end{tabular}
\caption{Entropy change  $\Delta S_{\rm o\rightarrow fb}/k_B$ from the open to the fully bonded state for the three polymer types. The error associated to $\Delta S_{\rm o\rightarrow fb}/k_B$ is of the order of $10^{-1}$.
The last column reports the gyration radius $R_g^2 / \sigma^2$, where $\sigma$ is the unit of length, corresponding to the monomer diameter. \label{tbl:entropy}
}
\end{center}
\end{table}%

%The table also shows the gyration radius to highlight  the significant size reduction on going from identical to distinct reactive types.

\begin{figure}[htbp] %  figure placement: here, top, bottom, or page
   \centering
      \includegraphics[width=0.395\textwidth]{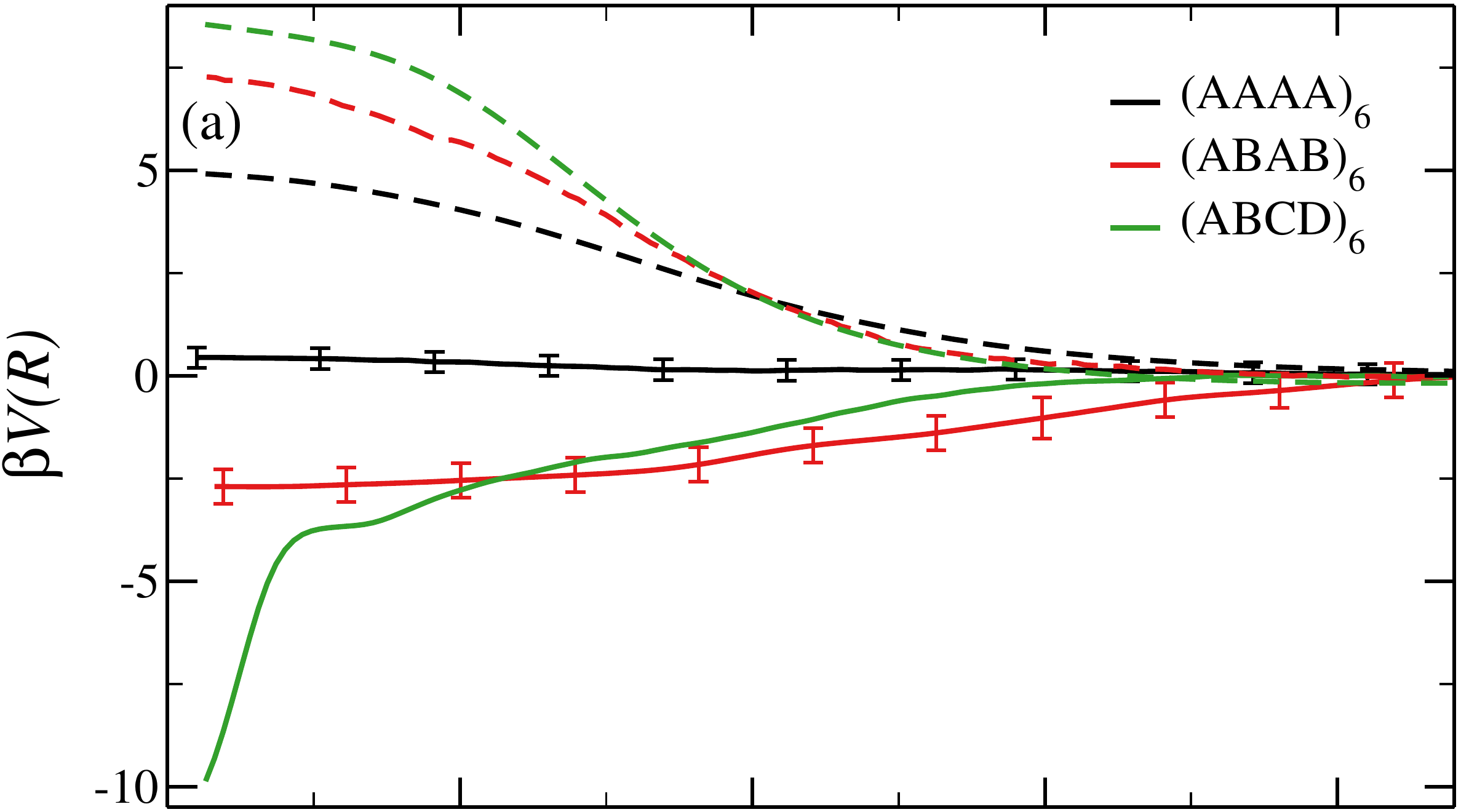} 
            \includegraphics[width=0.4\textwidth]{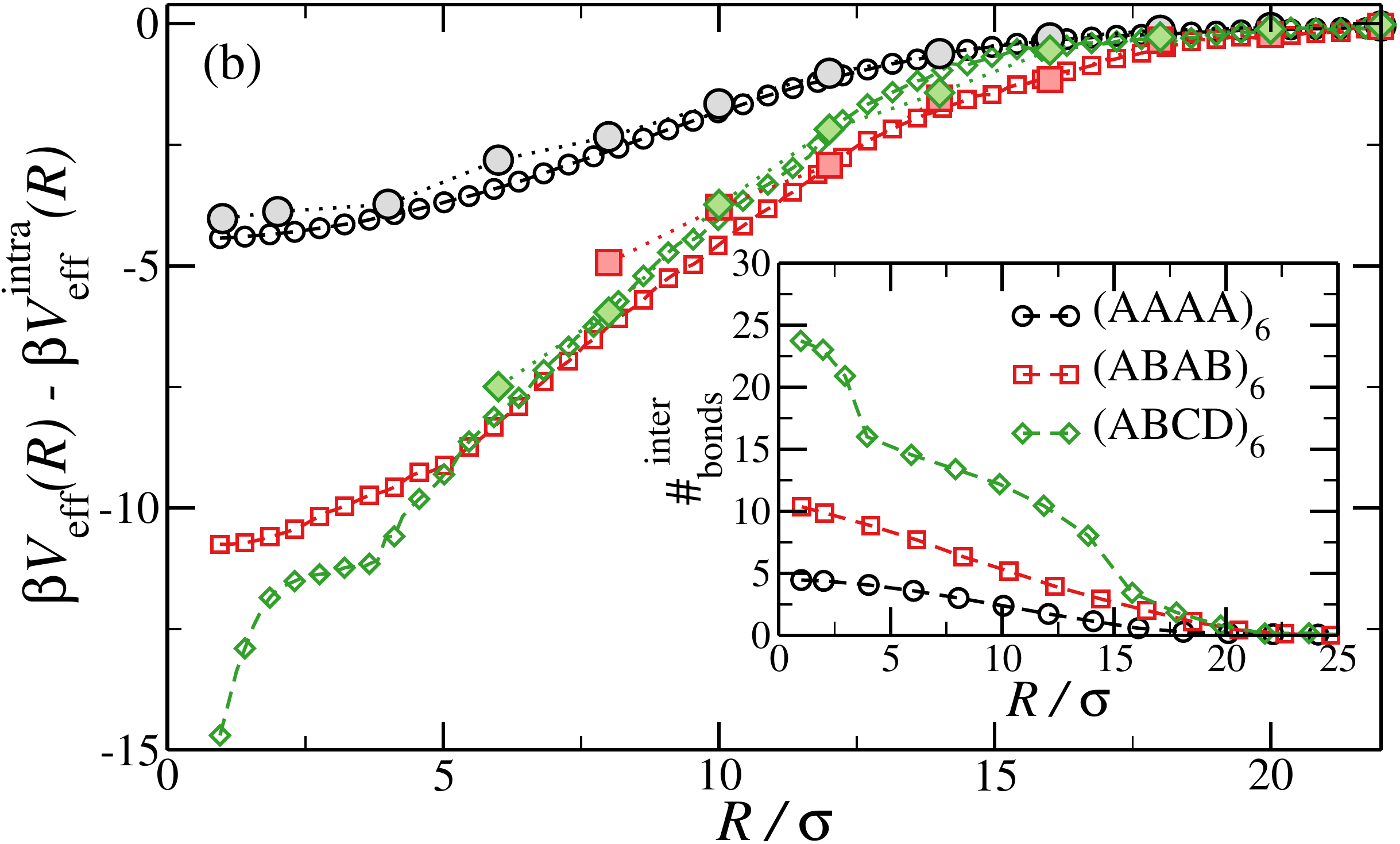} 
   \caption{(a) Effective potentials for the three polymers: $\beta V_{\rm eff}(R)$ and $\beta V_{\rm eff}^{\rm intra}(R)$ are shown with solid and dashed lines, respectively. The error bars are upper bounds estimated by splitting the data in two blocks and computing the absolute difference between the effective interactions in each block, divided by $\sqrt{2}$. (b) Attractive part of the potential  estimated as $\beta V_{\rm eff}(R)-\beta V_{\rm eff}^{\rm intra}(R)$ (open symbols) and as $\ln p(0)$ (see Eq.~\ref{eq:p0}, filled symbols). The inset shows the average number of inter-polymer bonds.  Note that the abrupt decrease of $\beta V_{\rm eff}(R)$  for very short $R$ in the (ABCD)$_6$ case originates from {\it zipping} of the bonds.
   }
   \label{fig:betaV1}
\end{figure}

{\bf Effective Potential (numerical evaluation):} 
To evaluate the  strength of the  entropic contributions we compute
(as described in Section~S1-C)
 $\beta V_{\rm eff}(R)$ for the three polymers. Note that we simulate under conditions that allow for bond breaking (such that the bond-swapping mechanism is active), but only configurations in which all possible bonds are formed are included in the statistical average. For each of the three models 
we also evaluate the potential, $\beta V_{\rm eff}^{\rm intra}(R)$ where only intra-bonds are allowed.
Results are shown in Fig.~\ref{fig:betaV1}.

Consistent with mean-field theory~\cite{semenov1998thermoreversible} and recent simulations~\cite{paciolla2021validity},
%The striking result of Fig.~\ref{fig:betaV1}(a) is that 
in the (AAAA)$_6$ case, despite the 
smaller $\beta V_{\rm eff}^{\rm intra}(R)$, $\beta V_{\rm eff}(R)$ is always positive and close to zero for all $R$, indicating that there is no net attraction between the polymers: The entropic attraction almost completely compensates the entropic repulsion.  
Differently and strikingly, in the other two cases, 
$\beta V_{\rm eff}(R)$  is strongly attractive,  suggesting the possibility of a phase separation.  
%Since, as discussed previously,  in principle the  combinatorial contribution is comparable in all three cases (Eq.~\ref{eq:sconfapprox}-\ref{eq:sconfapprox2}), the significant difference between the effective interactions must originate in the different $S_{\rm conf}$ contribution. 
Thus an appropriate design of the reactive monomer types 
 can be used to control the resulting inter-polymer attraction. To confirm the enhanced inter-polymer binding,  the inset of Fig.~\ref{fig:betaV1}(b) shows the number of inter-polymer bonds for the three cases.  In the (AAAA)$_6$ 
case  only a limited number of inter-polymer bonds are formed, even when the 
relative distance between the two polymers approaches zero. The conformational entropic gain of opening (two) intra-polymer bonds to form  (two) inter-polymer bonds does not sufficiently compensate  the entropic repulsion.
 The difference $\beta V_{\rm eff}(R)-\beta V_{\rm eff}^{\rm intra}(R)$ provides a measure of the total entropic attraction between two polymers (sum of the combinatorial and of the conformational terms) and it is shown in Fig.~\ref{fig:betaV1}(b). 
 %The different behavior of  $V_{\rm eff}(R)-V_{\rm eff}^{\rm intra}(R)$  for the three polymer cases confirms the relevant role played by the conformational contribution. 
% As we have discussed above, if $S_{\rm conf}$ was negligible or similar in the three SCNP, the attractive contribution would be comparable for all three  topologies.  Instead, 
The contribution $\beta V_{\rm eff}(R)-\beta V_{\rm eff}^{\rm intra}(R)$ for the (AAAA)$_6$ and for the (ABAB)$_6$ (and the (ABCD)$_6$) sequences is quite different, confirming the different role played by entropy for the three polymer cases. 

To support the numerical results for $\beta V_{\rm eff}(R)-\beta V_{\rm eff}^{\rm intra}(R)$ we estimate
the entropic attraction in an independent way. The partition function $Z$ of the two-polymer system at fixed relative distance, when both intra and inter bonds are possible, can be approximated as a sum over the number of inter-polymer bonds $\#_{b}$ of a specific type, from 0 to the maximum number of bonds $N_R$:

\begin{equation}
Z= \sum_{{0 \leq \#_{b} \leq N_R}\atop{\#_{b} \text{even number}}} \Omega_{\#_{b}}.
\label{eq:zeta}
\end{equation}
The fully bonded condition imposes only even numbers for $\#_{b}$. Here $\Omega_{\#_{b}}$ counts the number of micro-states available to the two chains when $\#_{b}$ inter-bonds are present.
The analogous expression, with the constraint of only intra-polymer bonds,
would include only the first term of the sum ($\#_{b}=0$) in Eq.~\ref{eq:zeta}.
Hence the entropic loss $\Delta S/k_B$ on going from inter and intra bonds 
to only intra bonds is

\begin{equation}
\frac{ \Delta S}{k_B} =  \ln \frac{\Omega_{0}}{Z} \equiv \ln p(0),
\label{eq:p0}
\end{equation}
\noindent
The quantity $\ln p(0)$, which provides a neat (and independent) measure of the entropic attractive contribution, is also reported in Fig.~\ref{fig:betaV1}(b) and favourably compares with 
$\beta V_{\rm eff}(R)-\beta V_{\rm eff}^{\rm intra}(R)$.

In Section~S2 we show
that the observed trends  are robust for changes in the polymer length (at fixed number of reactive monomers, by increasing the number of inert monomers) as well as for changes at fixed polymer length on changing the number of reactive monomers. 

\begin{figure}[h!]
\includegraphics[width=0.45\textwidth]{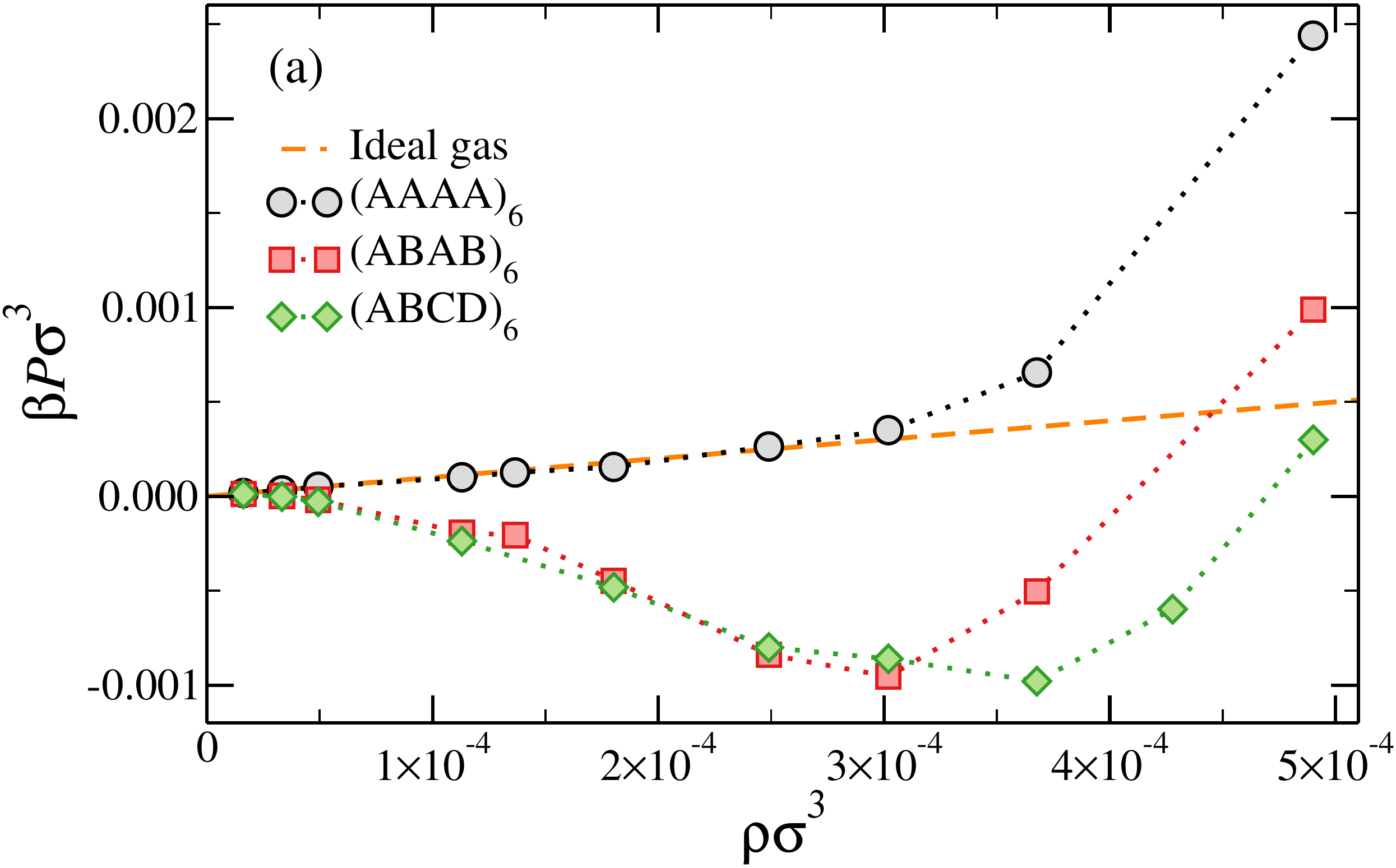}\\
\includegraphics[width=0.45\textwidth]{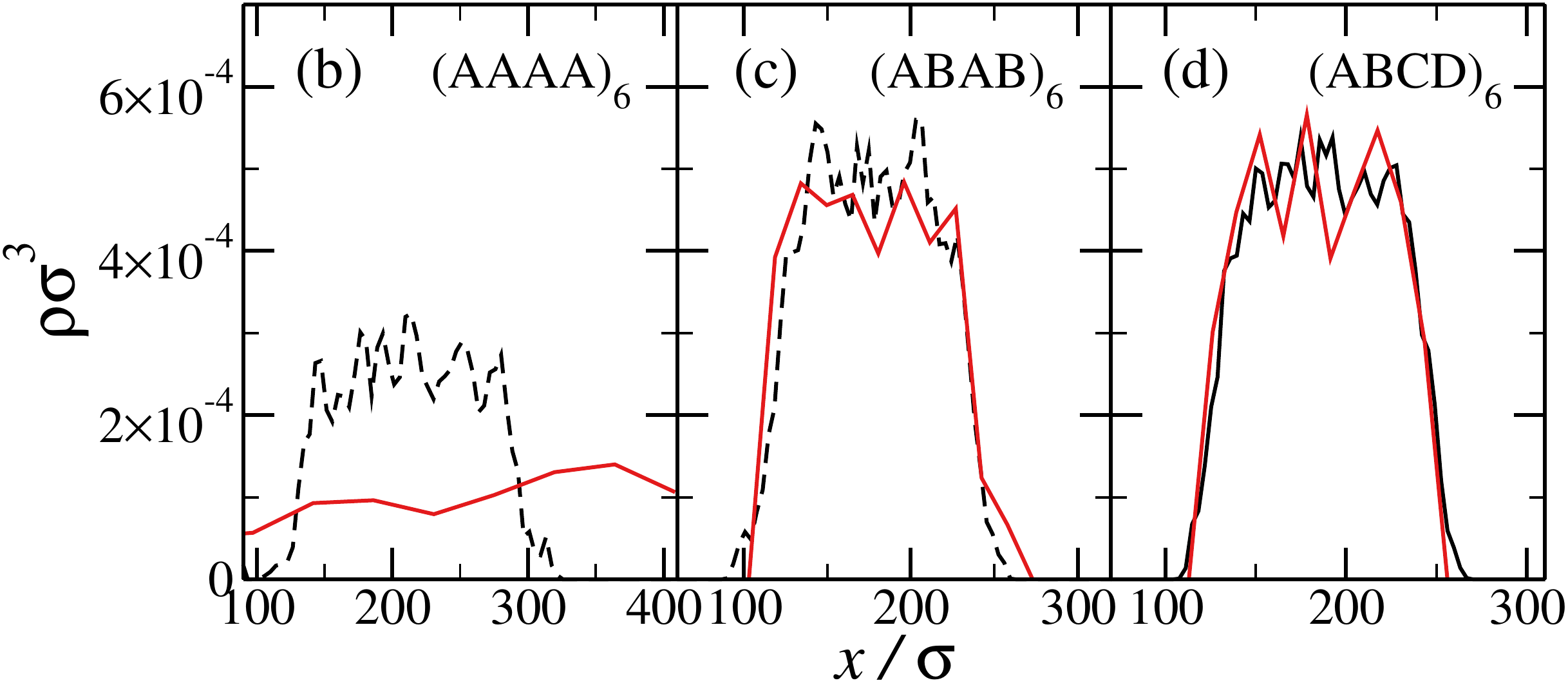}\\
\hspace{3.5em}\includegraphics[width=190px]{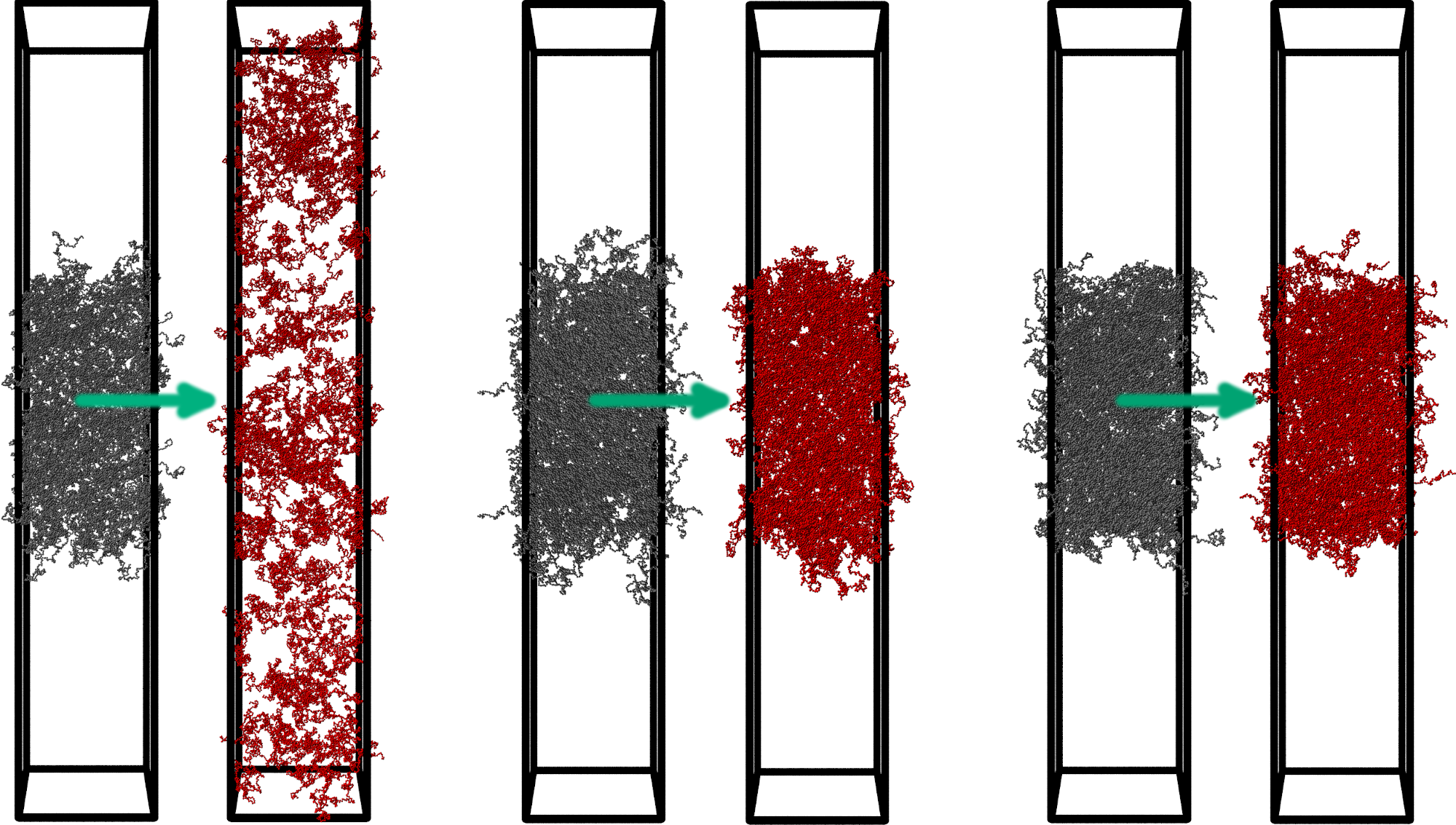}
\caption{(a) Equations of state of the three investigated systems (points) and of an ideal gas of chains (orange dashed line). The statistical uncertainty is smaller than the symbol size. (b-d) Density profiles along $x$ for the three investigated systems evaluated with direct coexistence simulations. We show the profiles of the density of the initial configuration (dashed black lines) and of the average density of the final state (solid red lines). Representative snaphosts of the initial (grey) and final (red) states are shown below each plot.
\label{fig:eos}}
\end{figure}

{\bf Phase Behavior:} To confirm that the entropic attraction  for the (ABAB)$_6$ and (ABCD)$_6$ polymers 
is sufficiently strong to condensate  a ``liquid''   from the ``gas'', we evaluate
 their equation of state  (Fig.~\ref{fig:eos}(a)), calculated as discussed in Section~S1-D. Coherently with the two-body effective interaction results, the (AAAA)$_6$  system behaves essentially as an ideal gas.
 Interestingly, this $\theta$-condition  originates from the
ability of the entropic attraction to essentially compensate 
the usual polymer repulsion.  By contrast, the pressure in both the (ABAB)$_6$ and (ABCD)$_6$ systems becomes quickly negative, suggesting the presence of a phase transition between two phases with significantly different polymer concentration.
As a  proof of the possible presence of a phase separation we perform  direct-coexistence simulations (Section~S1-E), by preparing a starting configuration composed by an inhomogeneous polymer concentration. Fig.~\ref{fig:eos}(b)  shows the density profiles of the initial  and final configurations for all three models. The density of the (AAAA)$_6$ system becomes homogeneous and the two interfaces that were present at time zero completely disappear. 
%, confirming that single-value behavior of the equation of state. 
By contrast, the   gas-liquid interfaces are stable in both the   (ABAB)$_6$ and (ABCD)$_6$ systems  over the course of the simulation. 
We find (not shown) that in the liquid phase at coexistence the system percolates and that
each polymer binds with $\approx 10$  other polymers for the (ABAB)$_6$ and (ABCD)$_6$ models.

%In order to double check that model B-24 does not phase separate while the other two do not we also run direct-coexistence simulations. We prepare the initial configuration by taking an equilibrium configuration, replicating it along the $x$ axis and increasing the resulting box size by a factor of $3$. We then run a long simulation and follow the evolution of the system.

%Finally  Fig.~\ref{fig:eos}(c) shows the concentration dependence of the connectivity of the system, defined as the average number of inter-polymer bonds per chain, $\langle N_n \rangle$. In the studied range for the (AAAA)$_6$ model $\langle N_n \rangle$ vanishes almost linearly with the concentration, confirming that a low $\rho$ the polymers preferentially form fully bonded isolated chains, in agreement with the effective potential findings.  By contrast, $\langle N_n \rangle$ in the (ABAB)$_6$ and (ABCD)$_6$ cases is larger than 4 for all but the lowest probed value of $\rho$, confirming the clustering process that signals the onset of phase separation.  Interestingly, the density dependence of $\langle N_n \rangle$ for $\rho \gtrsim 10^{-4}$ is roughly the same for all three models.

%\begin{figure}[h!]
%\includegraphics[width=0.45\textwidth]{dc_density.pdf}
%\caption{The average number of neighbours per chain as a function of density for the three models (panel \textbf{a}) and the associated probability distributions for selected densities (panels \textbf{b} and \textbf{c} for the B-12 and BC-12 models, respectively).
%\label{fig:dc_density}}
%\end{figure}

{\bf Concluding Remarks:}  In summary, we have demonstrated that
%that the collective behavior of single chain nano particles is dominated by entropy.
opposite to what previously expected,  a gas-liquid phase separation can appear in systems of SCNPs.
%we have provided evidence that by
 By simply alternating two different types of reactive monomers, it is possible to tune the  conformational entropy change on swapping intra to inter binding. As a result,  the strength of
 configurational and   combinatorial entropy can be harnessed to induce   
a fully entropic first-order phase transition, even with a low overall concentration of
reactive monomers.  
In view of the modern ability to force living cells to 
express SCNP, it is foreseeable to imagine --- guided by entropy --- a fine tuned control of the phase behavior of these particles in biologically relevant conditions~\cite{nakamura2018intracellular} as well as a mechanism to optimise multivalent binding~\cite{dubacheva2014superselective,liu2020combinatorial} in  ligand-receptors equilibria. 
At the same time, the design principles reported here may help achieving a better understanding of phase separation phenomena in cells, which are most often mediated by multivalent biomacromolecules~\cite{hyman2014liquid,brangwynne2015polymer,banani2017biomolecular}.  Such phenomena are commonly termed ``liquid-liquid''  transitions, since in both phases proteins are dispersed in an aqueous solvent --- the protein rich and poor phases being respectively the liquid and the gas in the present (implicit solvent) model.

{\bf Acknowledgement:}
 We acknowledge support from  MIUR PRIN 2017 (Project 2017Z55KCW) and the CINECA award under the ISCRA initiative, for the availability of high performance computing resources and support (Iscra B "AssoPoN"). We thank Angel Moreno for fruitful discussions.

\bibliography{final.bib}

 \newcommand{\noop}[1]{}
\begin{thebibliography}{54}%
\makeatletter
\providecommand \@ifxundefined [1]{%
 \@ifx{#1\undefined}
}%
\providecommand \@ifnum [1]{%
 \ifnum #1\expandafter \@firstoftwo
 \else \expandafter \@secondoftwo
 \fi
}%
\providecommand \@ifx [1]{%
 \ifx #1\expandafter \@firstoftwo
 \else \expandafter \@secondoftwo
 \fi
}%
\providecommand \natexlab [1]{#1}%
\providecommand \enquote  [1]{``#1''}%
\providecommand \bibnamefont  [1]{#1}%
\providecommand \bibfnamefont [1]{#1}%
\providecommand \citenamefont [1]{#1}%
\providecommand \href@noop [0]{\@secondoftwo}%
\providecommand \href [0]{\begingroup \@sanitize@url \@href}%
\providecommand \@href[1]{\@@startlink{#1}\@@href}%
\providecommand \@@href[1]{\endgroup#1\@@endlink}%
\providecommand \@sanitize@url [0]{\catcode `\\12\catcode `\$12\catcode
  `\&12\catcode `\#12\catcode `\^12\catcode `\_12\catcode `\%12\relax}%
\providecommand \@@startlink[1]{}%
\providecommand \@@endlink[0]{}%
\providecommand \url  [0]{\begingroup\@sanitize@url \@url }%
\providecommand \@url [1]{\endgroup\@href {#1}{\urlprefix }}%
\providecommand \urlprefix  [0]{URL }%
\providecommand \Eprint [0]{\href }%
\providecommand \doibase [0]{http://dx.doi.org/}%
\providecommand \selectlanguage [0]{\@gobble}%
\providecommand \bibinfo  [0]{\@secondoftwo}%
\providecommand \bibfield  [0]{\@secondoftwo}%
\providecommand \translation [1]{[#1]}%
\providecommand \BibitemOpen [0]{}%
\providecommand \bibitemStop [0]{}%
\providecommand \bibitemNoStop [0]{.\EOS\space}%
\providecommand \EOS [0]{\spacefactor3000\relax}%
\providecommand \BibitemShut  [1]{\csname bibitem#1\endcsname}%
\let\auto@bib@innerbib\@empty
%</preamble>
\bibitem [{\citenamefont {{Van der Waals}}(1873)}]{vdw}%
  \BibitemOpen
  \bibfield  {author} {\bibinfo {author} {\bibnamefont {{Van der Waals}}},\
  }\href@noop {} {Ph.D. thesis},\ \bibinfo  {school} {University of Leiden}
  (\bibinfo {year} {1873})\BibitemShut {NoStop}%
\bibitem [{\citenamefont {Asakura}\ and\ \citenamefont
  {Oosawa}(1954)}]{asakura1954interaction}%
  \BibitemOpen
  \bibfield  {author} {\bibinfo {author} {\bibfnamefont {Sho}\ \bibnamefont
  {Asakura}}\ and\ \bibinfo {author} {\bibfnamefont {Fumio}\ \bibnamefont
  {Oosawa}},\ }\bibfield  {title} {\enquote {\bibinfo {title} {{On interaction
  between two bodies immersed in a solution of macromolecules}},}\ }\href@noop
  {} {\bibfield  {journal} {\bibinfo  {journal} {The Journal of Chemical
  Physics}\ }\textbf {\bibinfo {volume} {22}},\ \bibinfo {pages} {1255--1256}
  (\bibinfo {year} {1954})}\BibitemShut {NoStop}%
\bibitem [{\citenamefont {Tuinier}\ and\ \citenamefont
  {Lekkerkerker}(2011)}]{tuinier2011colloids}%
  \BibitemOpen
  \bibfield  {author} {\bibinfo {author} {\bibfnamefont {Remco}\ \bibnamefont
  {Tuinier}}\ and\ \bibinfo {author} {\bibfnamefont {Henk~NW}\ \bibnamefont
  {Lekkerkerker}},\ }\href@noop {} {\emph {\bibinfo {title} {Colloids and the
  depletion interaction}}}\ (\bibinfo  {publisher} {Springer Netherlands},\
  \bibinfo {year} {2011})\BibitemShut {NoStop}%
\bibitem [{\citenamefont {Zilman}\ \emph {et~al.}(2003)\citenamefont {Zilman},
  \citenamefont {Kieffer}, \citenamefont {Molino}, \citenamefont {Porte},\ and\
  \citenamefont {Safran}}]{zilman2003entropic}%
  \BibitemOpen
  \bibfield  {author} {\bibinfo {author} {\bibfnamefont {A}~\bibnamefont
  {Zilman}}, \bibinfo {author} {\bibfnamefont {J}~\bibnamefont {Kieffer}},
  \bibinfo {author} {\bibfnamefont {F}~\bibnamefont {Molino}}, \bibinfo
  {author} {\bibfnamefont {G}~\bibnamefont {Porte}}, \ and\ \bibinfo {author}
  {\bibfnamefont {SA}~\bibnamefont {Safran}},\ }\bibfield  {title} {\enquote
  {\bibinfo {title} {Entropic phase separation in polymer-microemulsion
  networks},}\ }\href@noop {} {\bibfield  {journal} {\bibinfo  {journal}
  {Physical review letters}\ }\textbf {\bibinfo {volume} {91}},\ \bibinfo
  {pages} {015901} (\bibinfo {year} {2003})}\BibitemShut {NoStop}%
\bibitem [{\citenamefont {Lee}\ \emph {et~al.}(2019)\citenamefont {Lee},
  \citenamefont {Teich}, \citenamefont {Engel},\ and\ \citenamefont
  {Glotzer}}]{lee2019entropic}%
  \BibitemOpen
  \bibfield  {author} {\bibinfo {author} {\bibfnamefont {Sangmin}\ \bibnamefont
  {Lee}}, \bibinfo {author} {\bibfnamefont {Erin~G}\ \bibnamefont {Teich}},
  \bibinfo {author} {\bibfnamefont {Michael}\ \bibnamefont {Engel}}, \ and\
  \bibinfo {author} {\bibfnamefont {Sharon~C}\ \bibnamefont {Glotzer}},\
  }\bibfield  {title} {\enquote {\bibinfo {title} {Entropic colloidal
  crystallization pathways via fluid--fluid transitions and multidimensional
  prenucleation motifs},}\ }\href@noop {} {\bibfield  {journal} {\bibinfo
  {journal} {Proceedings of the National Academy of Sciences}\ }\textbf
  {\bibinfo {volume} {116}},\ \bibinfo {pages} {14843--14851} (\bibinfo {year}
  {2019})}\BibitemShut {NoStop}%
\bibitem [{\citenamefont {Oosawa}(2021)}]{doi:10.1063/5.0049350}%
  \BibitemOpen
  \bibfield  {author} {\bibinfo {author} {\bibfnamefont {Fumio}\ \bibnamefont
  {Oosawa}},\ }\bibfield  {title} {\enquote {\bibinfo {title} {The history of
  the birth of the asakura–oosawa theory},}\ }\href {\doibase
  10.1063/5.0049350} {\bibfield  {journal} {\bibinfo  {journal} {J. Chem.
  Phys.}\ }\textbf {\bibinfo {volume} {155}},\ \bibinfo {pages} {084104}
  (\bibinfo {year} {2021})}\BibitemShut {NoStop}%
\bibitem [{\citenamefont {Filali}\ \emph {et~al.}(2001)\citenamefont {Filali},
  \citenamefont {Ouazzani}, \citenamefont {Michel}, \citenamefont {Aznar},
  \citenamefont {Porte},\ and\ \citenamefont {Appell}}]{filali2001robust}%
  \BibitemOpen
  \bibfield  {author} {\bibinfo {author} {\bibfnamefont {Mohammed}\
  \bibnamefont {Filali}}, \bibinfo {author} {\bibfnamefont {Mohamed~Jamil}\
  \bibnamefont {Ouazzani}}, \bibinfo {author} {\bibfnamefont {Eric}\
  \bibnamefont {Michel}}, \bibinfo {author} {\bibfnamefont {Raymond}\
  \bibnamefont {Aznar}}, \bibinfo {author} {\bibfnamefont {Gr{\'e}goire}\
  \bibnamefont {Porte}}, \ and\ \bibinfo {author} {\bibfnamefont {Jacqueline}\
  \bibnamefont {Appell}},\ }\bibfield  {title} {\enquote {\bibinfo {title}
  {Robust phase behavior of model transient networks},}\ }\href@noop {}
  {\bibfield  {journal} {\bibinfo  {journal} {The Journal of Physical Chemistry
  B}\ }\textbf {\bibinfo {volume} {105}},\ \bibinfo {pages} {10528--10535}
  (\bibinfo {year} {2001})}\BibitemShut {NoStop}%
\bibitem [{\citenamefont {Angioletti-Uberti}\ \emph {et~al.}(2012)\citenamefont
  {Angioletti-Uberti}, \citenamefont {Mognetti},\ and\ \citenamefont
  {Frenkel}}]{angioletti2012re}%
  \BibitemOpen
  \bibfield  {author} {\bibinfo {author} {\bibfnamefont {Stefano}\ \bibnamefont
  {Angioletti-Uberti}}, \bibinfo {author} {\bibfnamefont {Bortolo~M}\
  \bibnamefont {Mognetti}}, \ and\ \bibinfo {author} {\bibfnamefont {Daan}\
  \bibnamefont {Frenkel}},\ }\bibfield  {title} {\enquote {\bibinfo {title}
  {Re-entrant melting as a design principle for dna-coated colloids},}\
  }\href@noop {} {\bibfield  {journal} {\bibinfo  {journal} {Nature materials}\
  }\textbf {\bibinfo {volume} {11}},\ \bibinfo {pages} {518--522} (\bibinfo
  {year} {2012})}\BibitemShut {NoStop}%
\bibitem [{\citenamefont {Bachmann}\ \emph {et~al.}(2016)\citenamefont
  {Bachmann}, \citenamefont {Kotar}, \citenamefont {Parolini}, \citenamefont
  {{\v{S}}ari{\'c}}, \citenamefont {Cicuta}, \citenamefont {Di~Michele},\ and\
  \citenamefont {Mognetti}}]{bachmann2016melting}%
  \BibitemOpen
  \bibfield  {author} {\bibinfo {author} {\bibfnamefont {Stephan~Jan}\
  \bibnamefont {Bachmann}}, \bibinfo {author} {\bibfnamefont {Jurij}\
  \bibnamefont {Kotar}}, \bibinfo {author} {\bibfnamefont {Lucia}\ \bibnamefont
  {Parolini}}, \bibinfo {author} {\bibfnamefont {An{\dj}ela}\ \bibnamefont
  {{\v{S}}ari{\'c}}}, \bibinfo {author} {\bibfnamefont {Pietro}\ \bibnamefont
  {Cicuta}}, \bibinfo {author} {\bibfnamefont {Lorenzo}\ \bibnamefont
  {Di~Michele}}, \ and\ \bibinfo {author} {\bibfnamefont {Bortolo~Matteo}\
  \bibnamefont {Mognetti}},\ }\bibfield  {title} {\enquote {\bibinfo {title}
  {Melting transition in lipid vesicles functionalised by mobile dna
  linkers},}\ }\href@noop {} {\bibfield  {journal} {\bibinfo  {journal} {Soft
  Matter}\ }\textbf {\bibinfo {volume} {12}},\ \bibinfo {pages} {7804--7817}
  (\bibinfo {year} {2016})}\BibitemShut {NoStop}%
\bibitem [{\citenamefont {Sciortino}(2019)}]{sciortino2019entropy}%
  \BibitemOpen
  \bibfield  {author} {\bibinfo {author} {\bibfnamefont {Francesco}\
  \bibnamefont {Sciortino}},\ }\bibfield  {title} {\enquote {\bibinfo {title}
  {Entropy in self-assembly},}\ }\href@noop {} {\bibfield  {journal} {\bibinfo
  {journal} {La Rivista del Nuovo Cimento}\ }\textbf {\bibinfo {volume} {42}},\
  \bibinfo {pages} {511--548} (\bibinfo {year} {2019})}\BibitemShut {NoStop}%
\bibitem [{\citenamefont {Sciortino}\ \emph {et~al.}(2020)\citenamefont
  {Sciortino}, \citenamefont {Zhang}, \citenamefont {Gang},\ and\ \citenamefont
  {Kumar}}]{sciortino2020combinatorial}%
  \BibitemOpen
  \bibfield  {author} {\bibinfo {author} {\bibfnamefont {Francesco}\
  \bibnamefont {Sciortino}}, \bibinfo {author} {\bibfnamefont {Yugang}\
  \bibnamefont {Zhang}}, \bibinfo {author} {\bibfnamefont {Oleg}\ \bibnamefont
  {Gang}}, \ and\ \bibinfo {author} {\bibfnamefont {Sanat~K}\ \bibnamefont
  {Kumar}},\ }\bibfield  {title} {\enquote {\bibinfo {title}
  {Combinatorial-entropy-driven aggregation in dna-grafted nanoparticles},}\
  }\href@noop {} {\bibfield  {journal} {\bibinfo  {journal} {ACS nano}\
  }\textbf {\bibinfo {volume} {14}},\ \bibinfo {pages} {5628--5635} (\bibinfo
  {year} {2020})}\BibitemShut {NoStop}%
\bibitem [{\citenamefont {Smallenburg}\ and\ \citenamefont
  {Sciortino}(2013)}]{smallenburg2013liquids}%
  \BibitemOpen
  \bibfield  {author} {\bibinfo {author} {\bibfnamefont {Frank}\ \bibnamefont
  {Smallenburg}}\ and\ \bibinfo {author} {\bibfnamefont {Francesco}\
  \bibnamefont {Sciortino}},\ }\bibfield  {title} {\enquote {\bibinfo {title}
  {Liquids more stable than crystals in particles with limited valence and
  flexible bonds},}\ }\href@noop {} {\bibfield  {journal} {\bibinfo  {journal}
  {Nature Physics}\ }\textbf {\bibinfo {volume} {9}},\ \bibinfo {pages}
  {554--558} (\bibinfo {year} {2013})}\BibitemShut {NoStop}%
\bibitem [{\citenamefont {Kumar}\ and\ \citenamefont
  {Douglas}(2001)}]{kumar2001gelation}%
  \BibitemOpen
  \bibfield  {author} {\bibinfo {author} {\bibfnamefont {Sanat~K}\ \bibnamefont
  {Kumar}}\ and\ \bibinfo {author} {\bibfnamefont {Jack~F}\ \bibnamefont
  {Douglas}},\ }\bibfield  {title} {\enquote {\bibinfo {title} {Gelation in
  physically associating polymer solutions},}\ }\href@noop {} {\bibfield
  {journal} {\bibinfo  {journal} {Physical review letters}\ }\textbf {\bibinfo
  {volume} {87}},\ \bibinfo {pages} {188301} (\bibinfo {year}
  {2001})}\BibitemShut {NoStop}%
\bibitem [{\citenamefont {Seo}\ \emph {et~al.}(2008)\citenamefont {Seo},
  \citenamefont {Beck}, \citenamefont {Paulusse}, \citenamefont {Hawker},\ and\
  \citenamefont {Kim}}]{seo2008polymeric}%
  \BibitemOpen
  \bibfield  {author} {\bibinfo {author} {\bibfnamefont {Myungeun}\
  \bibnamefont {Seo}}, \bibinfo {author} {\bibfnamefont {Benjamin~J}\
  \bibnamefont {Beck}}, \bibinfo {author} {\bibfnamefont {Jos~MJ}\ \bibnamefont
  {Paulusse}}, \bibinfo {author} {\bibfnamefont {Craig~J}\ \bibnamefont
  {Hawker}}, \ and\ \bibinfo {author} {\bibfnamefont {Sang~Youl}\ \bibnamefont
  {Kim}},\ }\bibfield  {title} {\enquote {\bibinfo {title} {Polymeric
  nanoparticles via noncovalent cross-linking of linear chains},}\ }\href@noop
  {} {\bibfield  {journal} {\bibinfo  {journal} {Macromolecules}\ }\textbf
  {\bibinfo {volume} {41}},\ \bibinfo {pages} {6413--6418} (\bibinfo {year}
  {2008})}\BibitemShut {NoStop}%
\bibitem [{\citenamefont {Lyon}\ \emph {et~al.}(2015)\citenamefont {Lyon},
  \citenamefont {Prasher}, \citenamefont {Hanlon}, \citenamefont {Tuten},
  \citenamefont {Tooley}, \citenamefont {Frank},\ and\ \citenamefont
  {Berda}}]{lyon2015brief}%
  \BibitemOpen
  \bibfield  {author} {\bibinfo {author} {\bibfnamefont {Christopher~K}\
  \bibnamefont {Lyon}}, \bibinfo {author} {\bibfnamefont {Alka}\ \bibnamefont
  {Prasher}}, \bibinfo {author} {\bibfnamefont {Ashley~M}\ \bibnamefont
  {Hanlon}}, \bibinfo {author} {\bibfnamefont {Bryan~T}\ \bibnamefont {Tuten}},
  \bibinfo {author} {\bibfnamefont {Christian~A}\ \bibnamefont {Tooley}},
  \bibinfo {author} {\bibfnamefont {Peter~G}\ \bibnamefont {Frank}}, \ and\
  \bibinfo {author} {\bibfnamefont {Erik~B}\ \bibnamefont {Berda}},\ }\bibfield
   {title} {\enquote {\bibinfo {title} {A brief user's guide to single-chain
  nanoparticles},}\ }\href@noop {} {\bibfield  {journal} {\bibinfo  {journal}
  {Polymer Chemistry}\ }\textbf {\bibinfo {volume} {6}},\ \bibinfo {pages}
  {181--197} (\bibinfo {year} {2015})}\BibitemShut {NoStop}%
\bibitem [{\citenamefont {Arbe}\ \emph {et~al.}(2016)\citenamefont {Arbe},
  \citenamefont {Pomposo}, \citenamefont {Moreno}, \citenamefont {LoVerso},
  \citenamefont {Gonz{\'a}lez-Burgos}, \citenamefont {Asenjo-Sanz},
  \citenamefont {Iturrospe}, \citenamefont {Radulescu}, \citenamefont
  {Ivanova},\ and\ \citenamefont {Colmenero}}]{arbe2016structure}%
  \BibitemOpen
  \bibfield  {author} {\bibinfo {author} {\bibfnamefont {Arantxa}\ \bibnamefont
  {Arbe}}, \bibinfo {author} {\bibfnamefont {Jos{\'e}~A}\ \bibnamefont
  {Pomposo}}, \bibinfo {author} {\bibfnamefont {Angel~J}\ \bibnamefont
  {Moreno}}, \bibinfo {author} {\bibfnamefont {F}~\bibnamefont {LoVerso}},
  \bibinfo {author} {\bibfnamefont {M}~\bibnamefont {Gonz{\'a}lez-Burgos}},
  \bibinfo {author} {\bibfnamefont {I}~\bibnamefont {Asenjo-Sanz}}, \bibinfo
  {author} {\bibfnamefont {A}~\bibnamefont {Iturrospe}}, \bibinfo {author}
  {\bibfnamefont {A}~\bibnamefont {Radulescu}}, \bibinfo {author}
  {\bibfnamefont {O}~\bibnamefont {Ivanova}}, \ and\ \bibinfo {author}
  {\bibfnamefont {J}~\bibnamefont {Colmenero}},\ }\bibfield  {title} {\enquote
  {\bibinfo {title} {Structure and dynamics of single-chain nano-particles in
  solution},}\ }\href@noop {} {\bibfield  {journal} {\bibinfo  {journal}
  {Polymer}\ }\textbf {\bibinfo {volume} {105}},\ \bibinfo {pages} {532--544}
  (\bibinfo {year} {2016})}\BibitemShut {NoStop}%
\bibitem [{\citenamefont {Pomposo}(2017)}]{pomposo2017single}%
  \BibitemOpen
  \bibfield  {author} {\bibinfo {author} {\bibfnamefont {Jos{\'e}~A}\
  \bibnamefont {Pomposo}},\ }\href@noop {} {\emph {\bibinfo {title}
  {Single-Chain Polymer Nanoparticles: Synthesis, Characterization,
  Simulations, and Applications}}}\ (\bibinfo  {publisher} {John Wiley \&
  Sons},\ \bibinfo {year} {2017})\BibitemShut {NoStop}%
\bibitem [{\citenamefont {Whitaker}\ \emph {et~al.}(2013)\citenamefont
  {Whitaker}, \citenamefont {Mahon},\ and\ \citenamefont
  {Fulton}}]{whitaker2013thermoresponsive}%
  \BibitemOpen
  \bibfield  {author} {\bibinfo {author} {\bibfnamefont {Daniel~E}\
  \bibnamefont {Whitaker}}, \bibinfo {author} {\bibfnamefont {Clare~S}\
  \bibnamefont {Mahon}}, \ and\ \bibinfo {author} {\bibfnamefont {David~A}\
  \bibnamefont {Fulton}},\ }\bibfield  {title} {\enquote {\bibinfo {title}
  {Thermoresponsive dynamic covalent single-chain polymer nanoparticles
  reversibly transform into a hydrogel},}\ }\href@noop {} {\bibfield  {journal}
  {\bibinfo  {journal} {Angewandte Chemie}\ }\textbf {\bibinfo {volume}
  {125}},\ \bibinfo {pages} {990--993} (\bibinfo {year} {2013})}\BibitemShut
  {NoStop}%
\bibitem [{\citenamefont {Tang}\ \emph {et~al.}(2015)\citenamefont {Tang},
  \citenamefont {Wang},\ and\ \citenamefont {Olsen}}]{tang2015anomalous}%
  \BibitemOpen
  \bibfield  {author} {\bibinfo {author} {\bibfnamefont {Shengchang}\
  \bibnamefont {Tang}}, \bibinfo {author} {\bibfnamefont {Muzhou}\ \bibnamefont
  {Wang}}, \ and\ \bibinfo {author} {\bibfnamefont {Bradley~D}\ \bibnamefont
  {Olsen}},\ }\bibfield  {title} {\enquote {\bibinfo {title} {Anomalous
  self-diffusion and sticky rouse dynamics in associative protein hydrogels},}\
  }\href@noop {} {\bibfield  {journal} {\bibinfo  {journal} {Journal of the
  American Chemical Society}\ }\textbf {\bibinfo {volume} {137}},\ \bibinfo
  {pages} {3946--3957} (\bibinfo {year} {2015})}\BibitemShut {NoStop}%
\bibitem [{\citenamefont {Ghosh}\ and\ \citenamefont
  {Schweizer}(2020)}]{ghosh2020physical}%
  \BibitemOpen
  \bibfield  {author} {\bibinfo {author} {\bibfnamefont {Ashesh}\ \bibnamefont
  {Ghosh}}\ and\ \bibinfo {author} {\bibfnamefont {Kenneth~S}\ \bibnamefont
  {Schweizer}},\ }\bibfield  {title} {\enquote {\bibinfo {title} {Physical bond
  breaking in associating copolymer liquids},}\ }\href@noop {} {\bibfield
  {journal} {\bibinfo  {journal} {ACS Macro Letters}\ }\textbf {\bibinfo
  {volume} {10}},\ \bibinfo {pages} {122--128} (\bibinfo {year}
  {2020})}\BibitemShut {NoStop}%
\bibitem [{\citenamefont {Oyarz{\'u}n}\ and\ \citenamefont
  {Mognetti}(2018)}]{oyarzun2018programming}%
  \BibitemOpen
  \bibfield  {author} {\bibinfo {author} {\bibfnamefont {Bernardo}\
  \bibnamefont {Oyarz{\'u}n}}\ and\ \bibinfo {author} {\bibfnamefont
  {Bortolo~Matteo}\ \bibnamefont {Mognetti}},\ }\bibfield  {title} {\enquote
  {\bibinfo {title} {Programming configurational changes in systems of
  functionalised polymers using reversible intramolecular linkages},}\
  }\href@noop {} {\bibfield  {journal} {\bibinfo  {journal} {Molecular
  Physics}\ }\textbf {\bibinfo {volume} {116}},\ \bibinfo {pages} {2927--2941}
  (\bibinfo {year} {2018})}\BibitemShut {NoStop}%
\bibitem [{\citenamefont {Statt}\ \emph {et~al.}(2020)\citenamefont {Statt},
  \citenamefont {Casademunt}, \citenamefont {Brangwynne},\ and\ \citenamefont
  {Panagiotopoulos}}]{statt2020model}%
  \BibitemOpen
  \bibfield  {author} {\bibinfo {author} {\bibfnamefont {Antonia}\ \bibnamefont
  {Statt}}, \bibinfo {author} {\bibfnamefont {Helena}\ \bibnamefont
  {Casademunt}}, \bibinfo {author} {\bibfnamefont {Clifford~P}\ \bibnamefont
  {Brangwynne}}, \ and\ \bibinfo {author} {\bibfnamefont {Athanassios~Z}\
  \bibnamefont {Panagiotopoulos}},\ }\bibfield  {title} {\enquote {\bibinfo
  {title} {Model for disordered proteins with strongly sequence-dependent
  liquid phase behavior},}\ }\href@noop {} {\bibfield  {journal} {\bibinfo
  {journal} {The Journal of chemical physics}\ }\textbf {\bibinfo {volume}
  {152}},\ \bibinfo {pages} {075101} (\bibinfo {year} {2020})}\BibitemShut
  {NoStop}%
\bibitem [{\citenamefont {Ruff}\ \emph {et~al.}(2021)\citenamefont {Ruff},
  \citenamefont {Dar},\ and\ \citenamefont {Pappu}}]{ruff2021ligand}%
  \BibitemOpen
  \bibfield  {author} {\bibinfo {author} {\bibfnamefont {Kiersten~M}\
  \bibnamefont {Ruff}}, \bibinfo {author} {\bibfnamefont {Furqan}\ \bibnamefont
  {Dar}}, \ and\ \bibinfo {author} {\bibfnamefont {Rohit~V}\ \bibnamefont
  {Pappu}},\ }\bibfield  {title} {\enquote {\bibinfo {title} {Ligand effects on
  phase separation of multivalent macromolecules},}\ }\href@noop {} {\bibfield
  {journal} {\bibinfo  {journal} {Proceedings of the National Academy of
  Sciences}\ }\textbf {\bibinfo {volume} {118}} (\bibinfo {year}
  {2021})}\BibitemShut {NoStop}%
\bibitem [{\citenamefont {Lichtinger}\ \emph {et~al.}(2021)\citenamefont
  {Lichtinger}, \citenamefont {Garaizar}, \citenamefont {Collepardo-Guevara},\
  and\ \citenamefont {Reinhardt}}]{lichtinger2021targeted}%
  \BibitemOpen
  \bibfield  {author} {\bibinfo {author} {\bibfnamefont {Simon~M}\ \bibnamefont
  {Lichtinger}}, \bibinfo {author} {\bibfnamefont {Adiran}\ \bibnamefont
  {Garaizar}}, \bibinfo {author} {\bibfnamefont {Rosana}\ \bibnamefont
  {Collepardo-Guevara}}, \ and\ \bibinfo {author} {\bibfnamefont {Aleks}\
  \bibnamefont {Reinhardt}},\ }\bibfield  {title} {\enquote {\bibinfo {title}
  {Targeted modulation of protein liquid--liquid phase separation by evolution
  of amino-acid sequence},}\ }\href@noop {} {\bibfield  {journal} {\bibinfo
  {journal} {PLOS Computational Biology}\ }\textbf {\bibinfo {volume} {17}},\
  \bibinfo {pages} {e1009328} (\bibinfo {year} {2021})}\BibitemShut {NoStop}%
\bibitem [{\citenamefont {Dignon}\ \emph {et~al.}(2018)\citenamefont {Dignon},
  \citenamefont {Zheng}, \citenamefont {Best}, \citenamefont {Kim},\ and\
  \citenamefont {Mittal}}]{dignon2018relation}%
  \BibitemOpen
  \bibfield  {author} {\bibinfo {author} {\bibfnamefont {Gregory~L}\
  \bibnamefont {Dignon}}, \bibinfo {author} {\bibfnamefont {Wenwei}\
  \bibnamefont {Zheng}}, \bibinfo {author} {\bibfnamefont {Robert~B}\
  \bibnamefont {Best}}, \bibinfo {author} {\bibfnamefont {Young~C}\
  \bibnamefont {Kim}}, \ and\ \bibinfo {author} {\bibfnamefont {Jeetain}\
  \bibnamefont {Mittal}},\ }\bibfield  {title} {\enquote {\bibinfo {title}
  {Relation between single-molecule properties and phase behavior of
  intrinsically disordered proteins},}\ }\href@noop {} {\bibfield  {journal}
  {\bibinfo  {journal} {Proceedings of the National Academy of Sciences}\
  }\textbf {\bibinfo {volume} {115}},\ \bibinfo {pages} {9929--9934} (\bibinfo
  {year} {2018})}\BibitemShut {NoStop}%
\bibitem [{\citenamefont {Zumbro}\ \emph {et~al.}(2019)\citenamefont {Zumbro},
  \citenamefont {Witten},\ and\ \citenamefont
  {Alexander-Katz}}]{ZUMBRO2019892}%
  \BibitemOpen
  \bibfield  {author} {\bibinfo {author} {\bibfnamefont {Emiko}\ \bibnamefont
  {Zumbro}}, \bibinfo {author} {\bibfnamefont {Jacob}\ \bibnamefont {Witten}},
  \ and\ \bibinfo {author} {\bibfnamefont {Alfredo}\ \bibnamefont
  {Alexander-Katz}},\ }\bibfield  {title} {\enquote {\bibinfo {title}
  {Computational insights into avidity of polymeric multivalent binders},}\
  }\href {\doibase https://doi.org/10.1016/j.bpj.2019.07.026} {\bibfield
  {journal} {\bibinfo  {journal} {Biophysical Journal}\ }\textbf {\bibinfo
  {volume} {117}},\ \bibinfo {pages} {892--902} (\bibinfo {year}
  {2019})}\BibitemShut {NoStop}%
\bibitem [{\citenamefont {Nakamura}\ \emph {et~al.}(2018)\citenamefont
  {Nakamura}, \citenamefont {Lee}, \citenamefont {Afshar}, \citenamefont
  {Watanabe}, \citenamefont {Rho}, \citenamefont {Razavi}, \citenamefont
  {Suarez}, \citenamefont {Lin}, \citenamefont {Tanigawa}, \citenamefont
  {Huang} \emph {et~al.}}]{nakamura2018intracellular}%
  \BibitemOpen
  \bibfield  {author} {\bibinfo {author} {\bibfnamefont {Hideki}\ \bibnamefont
  {Nakamura}}, \bibinfo {author} {\bibfnamefont {Albert~A}\ \bibnamefont
  {Lee}}, \bibinfo {author} {\bibfnamefont {Ali~Sobhi}\ \bibnamefont {Afshar}},
  \bibinfo {author} {\bibfnamefont {Shigeki}\ \bibnamefont {Watanabe}},
  \bibinfo {author} {\bibfnamefont {Elmer}\ \bibnamefont {Rho}}, \bibinfo
  {author} {\bibfnamefont {Shiva}\ \bibnamefont {Razavi}}, \bibinfo {author}
  {\bibfnamefont {Allister}\ \bibnamefont {Suarez}}, \bibinfo {author}
  {\bibfnamefont {Yu-Chun}\ \bibnamefont {Lin}}, \bibinfo {author}
  {\bibfnamefont {Makoto}\ \bibnamefont {Tanigawa}}, \bibinfo {author}
  {\bibfnamefont {Brian}\ \bibnamefont {Huang}},  \emph {et~al.},\ }\bibfield
  {title} {\enquote {\bibinfo {title} {Intracellular production of hydrogels
  and synthetic rna granules by multivalent molecular interactions},}\
  }\href@noop {} {\bibfield  {journal} {\bibinfo  {journal} {Nature materials}\
  }\textbf {\bibinfo {volume} {17}},\ \bibinfo {pages} {79--89} (\bibinfo
  {year} {2018})}\BibitemShut {NoStop}%
\bibitem [{\citenamefont {Moreno}\ \emph {et~al.}(2017)\citenamefont {Moreno},
  \citenamefont {Bacova}, \citenamefont {Verso}, \citenamefont {Arbe},
  \citenamefont {Colmenero},\ and\ \citenamefont {Pomposo}}]{moreno2017effect}%
  \BibitemOpen
  \bibfield  {author} {\bibinfo {author} {\bibfnamefont {Angel~J}\ \bibnamefont
  {Moreno}}, \bibinfo {author} {\bibfnamefont {Petra}\ \bibnamefont {Bacova}},
  \bibinfo {author} {\bibfnamefont {Federica~Lo}\ \bibnamefont {Verso}},
  \bibinfo {author} {\bibfnamefont {Arantxa}\ \bibnamefont {Arbe}}, \bibinfo
  {author} {\bibfnamefont {Juan}\ \bibnamefont {Colmenero}}, \ and\ \bibinfo
  {author} {\bibfnamefont {Jos{\'e}~A}\ \bibnamefont {Pomposo}},\ }\bibfield
  {title} {\enquote {\bibinfo {title} {Effect of chain stiffness on the
  structure of single-chain polymer nanoparticles},}\ }\href@noop {} {\bibfield
   {journal} {\bibinfo  {journal} {Journal of Physics: Condensed Matter}\
  }\textbf {\bibinfo {volume} {30}},\ \bibinfo {pages} {034001} (\bibinfo
  {year} {2017})}\BibitemShut {NoStop}%
\bibitem [{\citenamefont {Formanek}\ \emph {et~al.}(2021)\citenamefont
  {Formanek}, \citenamefont {Rovigatti}, \citenamefont {Zaccarelli},
  \citenamefont {Sciortino},\ and\ \citenamefont {Moreno}}]{formanek2021gel}%
  \BibitemOpen
  \bibfield  {author} {\bibinfo {author} {\bibfnamefont {Maud}\ \bibnamefont
  {Formanek}}, \bibinfo {author} {\bibfnamefont {Lorenzo}\ \bibnamefont
  {Rovigatti}}, \bibinfo {author} {\bibfnamefont {Emanuela}\ \bibnamefont
  {Zaccarelli}}, \bibinfo {author} {\bibfnamefont {Francesco}\ \bibnamefont
  {Sciortino}}, \ and\ \bibinfo {author} {\bibfnamefont {Angel~J}\ \bibnamefont
  {Moreno}},\ }\bibfield  {title} {\enquote {\bibinfo {title} {Gel formation in
  reversibly cross-linking polymers},}\ }\href@noop {} {\bibfield  {journal}
  {\bibinfo  {journal} {Macromolecules}\ }\textbf {\bibinfo {volume} {54}},\
  \bibinfo {pages} {6613--6627} (\bibinfo {year} {2021})}\BibitemShut {NoStop}%
\bibitem [{\citenamefont {Semenov}\ and\ \citenamefont
  {Rubinstein}(1998)}]{semenov1998thermoreversible}%
  \BibitemOpen
  \bibfield  {author} {\bibinfo {author} {\bibfnamefont {Alexander~N}\
  \bibnamefont {Semenov}}\ and\ \bibinfo {author} {\bibfnamefont {Michael}\
  \bibnamefont {Rubinstein}},\ }\bibfield  {title} {\enquote {\bibinfo {title}
  {Thermoreversible gelation in solutions of associative polymers. 1.
  statics},}\ }\href@noop {} {\bibfield  {journal} {\bibinfo  {journal}
  {Macromolecules}\ }\textbf {\bibinfo {volume} {31}},\ \bibinfo {pages}
  {1373--1385} (\bibinfo {year} {1998})}\BibitemShut {NoStop}%
\bibitem [{\citenamefont {Flory}(1942)}]{flory1942constitution}%
  \BibitemOpen
  \bibfield  {author} {\bibinfo {author} {\bibfnamefont {Paul~J}\ \bibnamefont
  {Flory}},\ }\bibfield  {title} {\enquote {\bibinfo {title} {Constitution of
  three-dimensional polymers and the theory of gelation.}}\ }\href@noop {}
  {\bibfield  {journal} {\bibinfo  {journal} {The Journal of Physical
  Chemistry}\ }\textbf {\bibinfo {volume} {46}},\ \bibinfo {pages} {132--140}
  (\bibinfo {year} {1942})}\BibitemShut {NoStop}%
\bibitem [{\citenamefont {Harmon}\ \emph {et~al.}(2017)\citenamefont {Harmon},
  \citenamefont {Holehouse}, \citenamefont {Rosen},\ and\ \citenamefont
  {Pappu}}]{harmon2017intrinsically}%
  \BibitemOpen
  \bibfield  {author} {\bibinfo {author} {\bibfnamefont {Tyler~S}\ \bibnamefont
  {Harmon}}, \bibinfo {author} {\bibfnamefont {Alex~S}\ \bibnamefont
  {Holehouse}}, \bibinfo {author} {\bibfnamefont {Michael~K}\ \bibnamefont
  {Rosen}}, \ and\ \bibinfo {author} {\bibfnamefont {Rohit~V}\ \bibnamefont
  {Pappu}},\ }\bibfield  {title} {\enquote {\bibinfo {title} {Intrinsically
  disordered linkers determine the interplay between phase separation and
  gelation in multivalent proteins},}\ }\href@noop {} {\bibfield  {journal}
  {\bibinfo  {journal} {elife}\ }\textbf {\bibinfo {volume} {6}},\ \bibinfo
  {pages} {e30294} (\bibinfo {year} {2017})}\BibitemShut {NoStop}%
\bibitem [{\citenamefont {Kremer}\ and\ \citenamefont
  {Grest}(1990)}]{kremer1990dynamics}%
  \BibitemOpen
  \bibfield  {author} {\bibinfo {author} {\bibfnamefont {Kurt}\ \bibnamefont
  {Kremer}}\ and\ \bibinfo {author} {\bibfnamefont {Gary~S}\ \bibnamefont
  {Grest}},\ }\bibfield  {title} {\enquote {\bibinfo {title} {{Dynamics of
  entangled linear polymer melts: A molecular-dynamics simulation}},}\
  }\href@noop {} {\bibfield  {journal} {\bibinfo  {journal} {The Journal of
  Chemical Physics}\ }\textbf {\bibinfo {volume} {92}},\ \bibinfo {pages}
  {5057--5086} (\bibinfo {year} {1990})}\BibitemShut {NoStop}%
\bibitem [{\citenamefont {Sciortino}(2017)}]{sciortino2017three}%
  \BibitemOpen
  \bibfield  {author} {\bibinfo {author} {\bibfnamefont {Francesco}\
  \bibnamefont {Sciortino}},\ }\bibfield  {title} {\enquote {\bibinfo {title}
  {Three-body potential for simulating bond swaps in molecular dynamics},}\
  }\href@noop {} {\bibfield  {journal} {\bibinfo  {journal} {The European
  Physical Journal E}\ }\textbf {\bibinfo {volume} {40}},\ \bibinfo {pages} {3}
  (\bibinfo {year} {2017})}\BibitemShut {NoStop}%
\bibitem [{\citenamefont {Gnan}\ \emph {et~al.}(2017)\citenamefont {Gnan},
  \citenamefont {Rovigatti}, \citenamefont {Bergman},\ and\ \citenamefont
  {Zaccarelli}}]{gnan2017silico}%
  \BibitemOpen
  \bibfield  {author} {\bibinfo {author} {\bibfnamefont {Nicoletta}\
  \bibnamefont {Gnan}}, \bibinfo {author} {\bibfnamefont {Lorenzo}\
  \bibnamefont {Rovigatti}}, \bibinfo {author} {\bibfnamefont {Maxime}\
  \bibnamefont {Bergman}}, \ and\ \bibinfo {author} {\bibfnamefont {Emanuela}\
  \bibnamefont {Zaccarelli}},\ }\bibfield  {title} {\enquote {\bibinfo {title}
  {In silico synthesis of microgel particles},}\ }\href@noop {} {\bibfield
  {journal} {\bibinfo  {journal} {Macromolecules}\ }\textbf {\bibinfo {volume}
  {50}},\ \bibinfo {pages} {8777--8786} (\bibinfo {year} {2017})}\BibitemShut
  {NoStop}%
\bibitem [{\citenamefont {Rovigatti}\ \emph {et~al.}(2018)\citenamefont
  {Rovigatti}, \citenamefont {Nava}, \citenamefont {Bellini},\ and\
  \citenamefont {Sciortino}}]{rovigatti2018self}%
  \BibitemOpen
  \bibfield  {author} {\bibinfo {author} {\bibfnamefont {Lorenzo}\ \bibnamefont
  {Rovigatti}}, \bibinfo {author} {\bibfnamefont {Giovanni}\ \bibnamefont
  {Nava}}, \bibinfo {author} {\bibfnamefont {Tommaso}\ \bibnamefont {Bellini}},
  \ and\ \bibinfo {author} {\bibfnamefont {Francesco}\ \bibnamefont
  {Sciortino}},\ }\bibfield  {title} {\enquote {\bibinfo {title} {Self-dynamics
  and collective swap-driven dynamics in a particle model for vitrimers},}\
  }\href@noop {} {\bibfield  {journal} {\bibinfo  {journal} {Macromolecules}\
  }\textbf {\bibinfo {volume} {51}},\ \bibinfo {pages} {1232--1241} (\bibinfo
  {year} {2018})}\BibitemShut {NoStop}%
\bibitem [{\citenamefont {Ciarella}\ \emph {et~al.}(2018)\citenamefont
  {Ciarella}, \citenamefont {Sciortino},\ and\ \citenamefont
  {Ellenbroek}}]{ciarella2018dynamics}%
  \BibitemOpen
  \bibfield  {author} {\bibinfo {author} {\bibfnamefont {Simone}\ \bibnamefont
  {Ciarella}}, \bibinfo {author} {\bibfnamefont {Francesco}\ \bibnamefont
  {Sciortino}}, \ and\ \bibinfo {author} {\bibfnamefont {Wouter~G}\
  \bibnamefont {Ellenbroek}},\ }\bibfield  {title} {\enquote {\bibinfo {title}
  {Dynamics of vitrimers: Defects as a highway to stress relaxation},}\
  }\href@noop {} {\bibfield  {journal} {\bibinfo  {journal} {Physical review
  letters}\ }\textbf {\bibinfo {volume} {121}},\ \bibinfo {pages} {058003}
  (\bibinfo {year} {2018})}\BibitemShut {NoStop}%
\bibitem [{\citenamefont {Sorichetti}\ \emph {et~al.}(2021)\citenamefont
  {Sorichetti}, \citenamefont {Ninarello}, \citenamefont {Ruiz-Franco},
  \citenamefont {Hugouvieux}, \citenamefont {Kob}, \citenamefont {Zaccarelli},\
  and\ \citenamefont {Rovigatti}}]{sorichetti2021effect}%
  \BibitemOpen
  \bibfield  {author} {\bibinfo {author} {\bibfnamefont {Valerio}\ \bibnamefont
  {Sorichetti}}, \bibinfo {author} {\bibfnamefont {Andrea}\ \bibnamefont
  {Ninarello}}, \bibinfo {author} {\bibfnamefont {Jos{\'e}~M}\ \bibnamefont
  {Ruiz-Franco}}, \bibinfo {author} {\bibfnamefont {Virginie}\ \bibnamefont
  {Hugouvieux}}, \bibinfo {author} {\bibfnamefont {Walter}\ \bibnamefont
  {Kob}}, \bibinfo {author} {\bibfnamefont {Emanuela}\ \bibnamefont
  {Zaccarelli}}, \ and\ \bibinfo {author} {\bibfnamefont {Lorenzo}\
  \bibnamefont {Rovigatti}},\ }\bibfield  {title} {\enquote {\bibinfo {title}
  {Effect of chain polydispersity on the elasticity of disordered polymer
  networks},}\ }\href@noop {} {\bibfield  {journal} {\bibinfo  {journal}
  {Macromolecules}\ }\textbf {\bibinfo {volume} {54}},\ \bibinfo {pages}
  {3769--3779} (\bibinfo {year} {2021})}\BibitemShut {NoStop}%
\bibitem [{sm()}]{sm}%
  \BibitemOpen
  \href@noop {} {\enquote {\bibinfo {title} {See supplemental material at
  http://link.aps.org/supplemental/xxxx/xxxx for a detailed description of the
  methods and for results about additional polymers, including
  refs.~\cite{stillinger1985computer,ferrenberg1989optimized,russo2009reversible,formanek2017effects}},}\
  }\BibitemShut {NoStop}%
\bibitem [{\citenamefont {Bianchi}\ \emph {et~al.}(2006)\citenamefont
  {Bianchi}, \citenamefont {Largo}, \citenamefont {Tartaglia}, \citenamefont
  {Zaccarelli},\ and\ \citenamefont {Sciortino}}]{bianchi2006phase}%
  \BibitemOpen
  \bibfield  {author} {\bibinfo {author} {\bibfnamefont {Emanuela}\
  \bibnamefont {Bianchi}}, \bibinfo {author} {\bibfnamefont {Julio}\
  \bibnamefont {Largo}}, \bibinfo {author} {\bibfnamefont {Piero}\ \bibnamefont
  {Tartaglia}}, \bibinfo {author} {\bibfnamefont {Emanuela}\ \bibnamefont
  {Zaccarelli}}, \ and\ \bibinfo {author} {\bibfnamefont {Francesco}\
  \bibnamefont {Sciortino}},\ }\bibfield  {title} {\enquote {\bibinfo {title}
  {Phase diagram of patchy colloids: Towards empty liquids},}\ }\href@noop {}
  {\bibfield  {journal} {\bibinfo  {journal} {Physical Review Letters}\
  }\textbf {\bibinfo {volume} {97}},\ \bibinfo {pages} {168301} (\bibinfo
  {year} {2006})}\BibitemShut {NoStop}%
\bibitem [{\citenamefont {Sciortino}\ and\ \citenamefont
  {Zaccarelli}(2017)}]{sciortino2017equilibrium}%
  \BibitemOpen
  \bibfield  {author} {\bibinfo {author} {\bibfnamefont {Francesco}\
  \bibnamefont {Sciortino}}\ and\ \bibinfo {author} {\bibfnamefont {Emanuela}\
  \bibnamefont {Zaccarelli}},\ }\bibfield  {title} {\enquote {\bibinfo {title}
  {Equilibrium gels of limited valence colloids},}\ }\href@noop {} {\bibfield
  {journal} {\bibinfo  {journal} {Current opinion in colloid \& interface
  science}\ }\textbf {\bibinfo {volume} {30}},\ \bibinfo {pages} {90--96}
  (\bibinfo {year} {2017})}\BibitemShut {NoStop}%
\bibitem [{\citenamefont {Grosberg}\ \emph {et~al.}(1982)\citenamefont
  {Grosberg}, \citenamefont {Khalatur},\ and\ \citenamefont
  {Khokhlov}}]{grosberg1982polymeric}%
  \BibitemOpen
  \bibfield  {author} {\bibinfo {author} {\bibfnamefont {Alexander~Yu}\
  \bibnamefont {Grosberg}}, \bibinfo {author} {\bibfnamefont {Pavel~G}\
  \bibnamefont {Khalatur}}, \ and\ \bibinfo {author} {\bibfnamefont {Alexei~R}\
  \bibnamefont {Khokhlov}},\ }\bibfield  {title} {\enquote {\bibinfo {title}
  {Polymeric coils with excluded volume in dilute solution: The invalidity of
  the model of impenetrable spheres and the influence of excluded volume on the
  rates of diffusion-controlled intermacromolecular reactions},}\ }\href@noop
  {} {\bibfield  {journal} {\bibinfo  {journal} {Die Makromolekulare Chemie,
  Rapid Communications}\ }\textbf {\bibinfo {volume} {3}},\ \bibinfo {pages}
  {709--713} (\bibinfo {year} {1982})}\BibitemShut {NoStop}%
\bibitem [{\citenamefont {Bolhuis}\ \emph {et~al.}(2001)\citenamefont
  {Bolhuis}, \citenamefont {Louis}, \citenamefont {Hansen},\ and\ \citenamefont
  {Meijer}}]{bolhuis2001accurate}%
  \BibitemOpen
  \bibfield  {author} {\bibinfo {author} {\bibfnamefont {PG}~\bibnamefont
  {Bolhuis}}, \bibinfo {author} {\bibfnamefont {AA}~\bibnamefont {Louis}},
  \bibinfo {author} {\bibfnamefont {JP}~\bibnamefont {Hansen}}, \ and\ \bibinfo
  {author} {\bibfnamefont {EJ}~\bibnamefont {Meijer}},\ }\bibfield  {title}
  {\enquote {\bibinfo {title} {Accurate effective pair potentials for polymer
  solutions},}\ }\href@noop {} {\bibfield  {journal} {\bibinfo  {journal} {The
  Journal of Chemical Physics}\ }\textbf {\bibinfo {volume} {114}},\ \bibinfo
  {pages} {4296--4311} (\bibinfo {year} {2001})}\BibitemShut {NoStop}%
\bibitem [{\citenamefont {Likos}(2001)}]{likos2001effective}%
  \BibitemOpen
  \bibfield  {author} {\bibinfo {author} {\bibfnamefont {Christos~N}\
  \bibnamefont {Likos}},\ }\bibfield  {title} {\enquote {\bibinfo {title}
  {{Effective interactions in soft condensed matter physics}},}\ }\href@noop {}
  {\bibfield  {journal} {\bibinfo  {journal} {Physics Reports}\ }\textbf
  {\bibinfo {volume} {348}},\ \bibinfo {pages} {267--439} (\bibinfo {year}
  {2001})}\BibitemShut {NoStop}%
\bibitem [{\citenamefont {Paciolla}\ \emph {et~al.}(2021)\citenamefont
  {Paciolla}, \citenamefont {Likos},\ and\ \citenamefont
  {Moreno}}]{paciolla2021validity}%
  \BibitemOpen
  \bibfield  {author} {\bibinfo {author} {\bibfnamefont {Mariarita}\
  \bibnamefont {Paciolla}}, \bibinfo {author} {\bibfnamefont {Christos~N}\
  \bibnamefont {Likos}}, \ and\ \bibinfo {author} {\bibfnamefont {Angel~J}\
  \bibnamefont {Moreno}},\ }\bibfield  {title} {\enquote {\bibinfo {title} {On
  the validity of effective potentials in crowded solutions of linear and ring
  polymers with reversible bonds},}\ }\href@noop {} {\bibfield  {journal}
  {\bibinfo  {journal} {arXiv preprint arXiv:2112.13067}\ } (\bibinfo {year}
  {2021})}\BibitemShut {NoStop}%
\bibitem [{\citenamefont {Dubacheva}\ \emph {et~al.}(2014)\citenamefont
  {Dubacheva}, \citenamefont {Curk}, \citenamefont {Mognetti}, \citenamefont
  {Auz{\'e}ly-Velty}, \citenamefont {Frenkel},\ and\ \citenamefont
  {Richter}}]{dubacheva2014superselective}%
  \BibitemOpen
  \bibfield  {author} {\bibinfo {author} {\bibfnamefont {Galina~V}\
  \bibnamefont {Dubacheva}}, \bibinfo {author} {\bibfnamefont {Tine}\
  \bibnamefont {Curk}}, \bibinfo {author} {\bibfnamefont {Bortolo~M}\
  \bibnamefont {Mognetti}}, \bibinfo {author} {\bibfnamefont {Rachel}\
  \bibnamefont {Auz{\'e}ly-Velty}}, \bibinfo {author} {\bibfnamefont {Daan}\
  \bibnamefont {Frenkel}}, \ and\ \bibinfo {author} {\bibfnamefont {Ralf~P}\
  \bibnamefont {Richter}},\ }\bibfield  {title} {\enquote {\bibinfo {title}
  {Superselective targeting using multivalent polymers},}\ }\href@noop {}
  {\bibfield  {journal} {\bibinfo  {journal} {Journal of the American Chemical
  Society}\ }\textbf {\bibinfo {volume} {136}},\ \bibinfo {pages} {1722--1725}
  (\bibinfo {year} {2014})}\BibitemShut {NoStop}%
\bibitem [{\citenamefont {Liu}\ \emph {et~al.}(2020)\citenamefont {Liu},
  \citenamefont {Apriceno}, \citenamefont {Sipin}, \citenamefont {Scarpa},
  \citenamefont {Rodriguez-Arco}, \citenamefont {Poma}, \citenamefont
  {Marchello}, \citenamefont {Battaglia},\ and\ \citenamefont
  {Angioletti-Uberti}}]{liu2020combinatorial}%
  \BibitemOpen
  \bibfield  {author} {\bibinfo {author} {\bibfnamefont {Meng}\ \bibnamefont
  {Liu}}, \bibinfo {author} {\bibfnamefont {Azzurra}\ \bibnamefont {Apriceno}},
  \bibinfo {author} {\bibfnamefont {Miguel}\ \bibnamefont {Sipin}}, \bibinfo
  {author} {\bibfnamefont {Edoardo}\ \bibnamefont {Scarpa}}, \bibinfo {author}
  {\bibfnamefont {Laura}\ \bibnamefont {Rodriguez-Arco}}, \bibinfo {author}
  {\bibfnamefont {Alessandro}\ \bibnamefont {Poma}}, \bibinfo {author}
  {\bibfnamefont {Gabriele}\ \bibnamefont {Marchello}}, \bibinfo {author}
  {\bibfnamefont {Giuseppe}\ \bibnamefont {Battaglia}}, \ and\ \bibinfo
  {author} {\bibfnamefont {Stefano}\ \bibnamefont {Angioletti-Uberti}},\
  }\bibfield  {title} {\enquote {\bibinfo {title} {Combinatorial entropy
  behaviour leads to range selective binding in ligand-receptor
  interactions},}\ }\href@noop {} {\bibfield  {journal} {\bibinfo  {journal}
  {Nature communications}\ }\textbf {\bibinfo {volume} {11}},\ \bibinfo {pages}
  {1--10} (\bibinfo {year} {2020})}\BibitemShut {NoStop}%
\bibitem [{\citenamefont {Hyman}\ \emph {et~al.}(2014)\citenamefont {Hyman},
  \citenamefont {Weber},\ and\ \citenamefont {J{\"u}licher}}]{hyman2014liquid}%
  \BibitemOpen
  \bibfield  {author} {\bibinfo {author} {\bibfnamefont {Anthony~A}\
  \bibnamefont {Hyman}}, \bibinfo {author} {\bibfnamefont {Christoph~A}\
  \bibnamefont {Weber}}, \ and\ \bibinfo {author} {\bibfnamefont {Frank}\
  \bibnamefont {J{\"u}licher}},\ }\bibfield  {title} {\enquote {\bibinfo
  {title} {Liquid-liquid phase separation in biology},}\ }\href@noop {}
  {\bibfield  {journal} {\bibinfo  {journal} {Annual review of cell and
  developmental biology}\ }\textbf {\bibinfo {volume} {30}},\ \bibinfo {pages}
  {39--58} (\bibinfo {year} {2014})}\BibitemShut {NoStop}%
\bibitem [{\citenamefont {Brangwynne}\ \emph {et~al.}(2015)\citenamefont
  {Brangwynne}, \citenamefont {Tompa},\ and\ \citenamefont
  {Pappu}}]{brangwynne2015polymer}%
  \BibitemOpen
  \bibfield  {author} {\bibinfo {author} {\bibfnamefont {Clifford~P}\
  \bibnamefont {Brangwynne}}, \bibinfo {author} {\bibfnamefont {Peter}\
  \bibnamefont {Tompa}}, \ and\ \bibinfo {author} {\bibfnamefont {Rohit~V}\
  \bibnamefont {Pappu}},\ }\bibfield  {title} {\enquote {\bibinfo {title}
  {Polymer physics of intracellular phase transitions},}\ }\href@noop {}
  {\bibfield  {journal} {\bibinfo  {journal} {Nature Physics}\ }\textbf
  {\bibinfo {volume} {11}},\ \bibinfo {pages} {899--904} (\bibinfo {year}
  {2015})}\BibitemShut {NoStop}%
\bibitem [{\citenamefont {Banani}\ \emph {et~al.}(2017)\citenamefont {Banani},
  \citenamefont {Lee}, \citenamefont {Hyman},\ and\ \citenamefont
  {Rosen}}]{banani2017biomolecular}%
  \BibitemOpen
  \bibfield  {author} {\bibinfo {author} {\bibfnamefont {Salman~F}\
  \bibnamefont {Banani}}, \bibinfo {author} {\bibfnamefont {Hyun~O}\
  \bibnamefont {Lee}}, \bibinfo {author} {\bibfnamefont {Anthony~A}\
  \bibnamefont {Hyman}}, \ and\ \bibinfo {author} {\bibfnamefont {Michael~K}\
  \bibnamefont {Rosen}},\ }\bibfield  {title} {\enquote {\bibinfo {title}
  {Biomolecular condensates: organizers of cellular biochemistry},}\
  }\href@noop {} {\bibfield  {journal} {\bibinfo  {journal} {Nature reviews
  Molecular cell biology}\ }\textbf {\bibinfo {volume} {18}},\ \bibinfo {pages}
  {285--298} (\bibinfo {year} {2017})}\BibitemShut {NoStop}%
\bibitem [{\citenamefont {Stillinger}\ and\ \citenamefont
  {Weber}(1985)}]{stillinger1985computer}%
  \BibitemOpen
  \bibfield  {author} {\bibinfo {author} {\bibfnamefont {Frank~H}\ \bibnamefont
  {Stillinger}}\ and\ \bibinfo {author} {\bibfnamefont {Thomas~A}\ \bibnamefont
  {Weber}},\ }\bibfield  {title} {\enquote {\bibinfo {title} {Computer
  simulation of local order in condensed phases of silicon},}\ }\href@noop {}
  {\bibfield  {journal} {\bibinfo  {journal} {Physical review B}\ }\textbf
  {\bibinfo {volume} {31}},\ \bibinfo {pages} {5262} (\bibinfo {year}
  {1985})}\BibitemShut {NoStop}%
\bibitem [{\citenamefont {Ferrenberg}\ and\ \citenamefont
  {Swendsen}(1989)}]{ferrenberg1989optimized}%
  \BibitemOpen
  \bibfield  {author} {\bibinfo {author} {\bibfnamefont {Alan~M}\ \bibnamefont
  {Ferrenberg}}\ and\ \bibinfo {author} {\bibfnamefont {Robert~H}\ \bibnamefont
  {Swendsen}},\ }\bibfield  {title} {\enquote {\bibinfo {title} {Optimized
  monte carlo data analysis},}\ }\href@noop {} {\bibfield  {journal} {\bibinfo
  {journal} {Computers in Physics}\ }\textbf {\bibinfo {volume} {3}},\ \bibinfo
  {pages} {101--104} (\bibinfo {year} {1989})}\BibitemShut {NoStop}%
\bibitem [{\citenamefont {Russo}\ \emph {et~al.}(2009)\citenamefont {Russo},
  \citenamefont {Tartaglia},\ and\ \citenamefont
  {Sciortino}}]{russo2009reversible}%
  \BibitemOpen
  \bibfield  {author} {\bibinfo {author} {\bibfnamefont {John}\ \bibnamefont
  {Russo}}, \bibinfo {author} {\bibfnamefont {Piero}\ \bibnamefont
  {Tartaglia}}, \ and\ \bibinfo {author} {\bibfnamefont {Francesco}\
  \bibnamefont {Sciortino}},\ }\bibfield  {title} {\enquote {\bibinfo {title}
  {Reversible gels of patchy particles: role of the valence},}\ }\href@noop {}
  {\bibfield  {journal} {\bibinfo  {journal} {The Journal of Chemical Physics}\
  }\textbf {\bibinfo {volume} {131}},\ \bibinfo {pages} {014504} (\bibinfo
  {year} {2009})}\BibitemShut {NoStop}%
\bibitem [{\citenamefont {Formanek}\ and\ \citenamefont
  {Moreno}(2017)}]{formanek2017effects}%
  \BibitemOpen
  \bibfield  {author} {\bibinfo {author} {\bibfnamefont {Maud}\ \bibnamefont
  {Formanek}}\ and\ \bibinfo {author} {\bibfnamefont {Angel~J}\ \bibnamefont
  {Moreno}},\ }\bibfield  {title} {\enquote {\bibinfo {title} {Effects of
  precursor topology and synthesis under crowding conditions on the structure
  of single-chain polymer nanoparticles},}\ }\href@noop {} {\bibfield
  {journal} {\bibinfo  {journal} {Soft Matter}\ }\textbf {\bibinfo {volume}
  {13}},\ \bibinfo {pages} {6430--6438} (\bibinfo {year} {2017})}\BibitemShut
  {NoStop}%
\end{thebibliography}%


 \newcommand{\noop}[1]{}
\begin{thebibliography}{10}%
\makeatletter
\providecommand \@ifxundefined [1]{%
 \@ifx{#1\undefined}
}%
\providecommand \@ifnum [1]{%
 \ifnum #1\expandafter \@firstoftwo
 \else \expandafter \@secondoftwo
 \fi
}%
\providecommand \@ifx [1]{%
 \ifx #1\expandafter \@firstoftwo
 \else \expandafter \@secondoftwo
 \fi
}%
\providecommand \natexlab [1]{#1}%
\providecommand \enquote  [1]{``#1''}%
\providecommand \bibnamefont  [1]{#1}%
\providecommand \bibfnamefont [1]{#1}%
\providecommand \citenamefont [1]{#1}%
\providecommand \href@noop [0]{\@secondoftwo}%
\providecommand \href [0]{\begingroup \@sanitize@url \@href}%
\providecommand \@href[1]{\@@startlink{#1}\@@href}%
\providecommand \@@href[1]{\endgroup#1\@@endlink}%
\providecommand \@sanitize@url [0]{\catcode `\\12\catcode `\$12\catcode
  `\&12\catcode `\#12\catcode `\^12\catcode `\_12\catcode `\%12\relax}%
\providecommand \@@startlink[1]{}%
\providecommand \@@endlink[0]{}%
\providecommand \url  [0]{\begingroup\@sanitize@url \@url }%
\providecommand \@url [1]{\endgroup\@href {#1}{\urlprefix }}%
\providecommand \urlprefix  [0]{URL }%
\providecommand \Eprint [0]{\href }%
\providecommand \doibase [0]{http://dx.doi.org/}%
\providecommand \selectlanguage [0]{\@gobble}%
\providecommand \bibinfo  [0]{\@secondoftwo}%
\providecommand \bibfield  [0]{\@secondoftwo}%
\providecommand \translation [1]{[#1]}%
\providecommand \BibitemOpen [0]{}%
\providecommand \bibitemStop [0]{}%
\providecommand \bibitemNoStop [0]{.\EOS\space}%
\providecommand \EOS [0]{\spacefactor3000\relax}%
\providecommand \BibitemShut  [1]{\csname bibitem#1\endcsname}%
\let\auto@bib@innerbib\@empty
%</preamble>
\bibitem [{\citenamefont {Kremer}\ and\ \citenamefont
  {Grest}(1990)}]{kremer1990dynamics}%
  \BibitemOpen
  \bibfield  {author} {\bibinfo {author} {\bibfnamefont {K.}~\bibnamefont
  {Kremer}}\ and\ \bibinfo {author} {\bibfnamefont {G.~S.}\ \bibnamefont
  {Grest}},\ }\href@noop {} {\bibfield  {journal} {\bibinfo  {journal} {The
  Journal of Chemical Physics}\ }\textbf {\bibinfo {volume} {92}},\ \bibinfo
  {pages} {5057} (\bibinfo {year} {1990})}\BibitemShut {NoStop}%
\bibitem [{\citenamefont {Stillinger}\ and\ \citenamefont
  {Weber}(1985)}]{stillinger1985computer}%
  \BibitemOpen
  \bibfield  {author} {\bibinfo {author} {\bibfnamefont {F.~H.}\ \bibnamefont
  {Stillinger}}\ and\ \bibinfo {author} {\bibfnamefont {T.~A.}\ \bibnamefont
  {Weber}},\ }\href@noop {} {\bibfield  {journal} {\bibinfo  {journal}
  {Physical review B}\ }\textbf {\bibinfo {volume} {31}},\ \bibinfo {pages}
  {5262} (\bibinfo {year} {1985})}\BibitemShut {NoStop}%
\bibitem [{\citenamefont {Sciortino}(2017)}]{sciortino2017three}%
  \BibitemOpen
  \bibfield  {author} {\bibinfo {author} {\bibfnamefont {F.}~\bibnamefont
  {Sciortino}},\ }\href@noop {} {\bibfield  {journal} {\bibinfo  {journal} {The
  European Physical Journal E}\ }\textbf {\bibinfo {volume} {40}},\ \bibinfo
  {pages} {3} (\bibinfo {year} {2017})}\BibitemShut {NoStop}%
\bibitem [{\citenamefont {Ferrenberg}\ and\ \citenamefont
  {Swendsen}(1989)}]{ferrenberg1989optimized}%
  \BibitemOpen
  \bibfield  {author} {\bibinfo {author} {\bibfnamefont {A.~M.}\ \bibnamefont
  {Ferrenberg}}\ and\ \bibinfo {author} {\bibfnamefont {R.~H.}\ \bibnamefont
  {Swendsen}},\ }\href@noop {} {\bibfield  {journal} {\bibinfo  {journal}
  {Computers in Physics}\ }\textbf {\bibinfo {volume} {3}},\ \bibinfo {pages}
  {101} (\bibinfo {year} {1989})}\BibitemShut {NoStop}%
\bibitem [{\citenamefont {Russo}\ \emph {et~al.}(2009)\citenamefont {Russo},
  \citenamefont {Tartaglia},\ and\ \citenamefont
  {Sciortino}}]{russo2009reversible}%
  \BibitemOpen
  \bibfield  {author} {\bibinfo {author} {\bibfnamefont {J.}~\bibnamefont
  {Russo}}, \bibinfo {author} {\bibfnamefont {P.}~\bibnamefont {Tartaglia}}, \
  and\ \bibinfo {author} {\bibfnamefont {F.}~\bibnamefont {Sciortino}},\
  }\href@noop {} {\bibfield  {journal} {\bibinfo  {journal} {The Journal of
  Chemical Physics}\ }\textbf {\bibinfo {volume} {131}},\ \bibinfo {pages}
  {014504} (\bibinfo {year} {2009})}\BibitemShut {NoStop}%
\bibitem [{\citenamefont {Moreno}\ \emph {et~al.}(2017)\citenamefont {Moreno},
  \citenamefont {Bacova}, \citenamefont {Verso}, \citenamefont {Arbe},
  \citenamefont {Colmenero},\ and\ \citenamefont {Pomposo}}]{moreno2017effect}%
  \BibitemOpen
  \bibfield  {author} {\bibinfo {author} {\bibfnamefont {A.~J.}\ \bibnamefont
  {Moreno}}, \bibinfo {author} {\bibfnamefont {P.}~\bibnamefont {Bacova}},
  \bibinfo {author} {\bibfnamefont {F.~L.}\ \bibnamefont {Verso}}, \bibinfo
  {author} {\bibfnamefont {A.}~\bibnamefont {Arbe}}, \bibinfo {author}
  {\bibfnamefont {J.}~\bibnamefont {Colmenero}}, \ and\ \bibinfo {author}
  {\bibfnamefont {J.~A.}\ \bibnamefont {Pomposo}},\ }\href@noop {} {\bibfield
  {journal} {\bibinfo  {journal} {Journal of Physics: Condensed Matter}\
  }\textbf {\bibinfo {volume} {30}},\ \bibinfo {pages} {034001} (\bibinfo
  {year} {2017})}\BibitemShut {NoStop}%
\bibitem [{\citenamefont {Formanek}\ and\ \citenamefont
  {Moreno}(2017)}]{formanek2017effects}%
  \BibitemOpen
  \bibfield  {author} {\bibinfo {author} {\bibfnamefont {M.}~\bibnamefont
  {Formanek}}\ and\ \bibinfo {author} {\bibfnamefont {A.~J.}\ \bibnamefont
  {Moreno}},\ }\href@noop {} {\bibfield  {journal} {\bibinfo  {journal} {Soft
  Matter}\ }\textbf {\bibinfo {volume} {13}},\ \bibinfo {pages} {6430}
  (\bibinfo {year} {2017})}\BibitemShut {NoStop}%
\bibitem [{\citenamefont {Semenov}\ and\ \citenamefont
  {Rubinstein}(1998)}]{semenov1998thermoreversible}%
  \BibitemOpen
  \bibfield  {author} {\bibinfo {author} {\bibfnamefont {A.~N.}\ \bibnamefont
  {Semenov}}\ and\ \bibinfo {author} {\bibfnamefont {M.}~\bibnamefont
  {Rubinstein}},\ }\href@noop {} {\bibfield  {journal} {\bibinfo  {journal}
  {Macromolecules}\ }\textbf {\bibinfo {volume} {31}},\ \bibinfo {pages} {1373}
  (\bibinfo {year} {1998})}\BibitemShut {NoStop}%
\bibitem [{\citenamefont {Zumbro}\ \emph {et~al.}(2019)\citenamefont {Zumbro},
  \citenamefont {Witten},\ and\ \citenamefont
  {Alexander-Katz}}]{ZUMBRO2019892}%
  \BibitemOpen
  \bibfield  {author} {\bibinfo {author} {\bibfnamefont {E.}~\bibnamefont
  {Zumbro}}, \bibinfo {author} {\bibfnamefont {J.}~\bibnamefont {Witten}}, \
  and\ \bibinfo {author} {\bibfnamefont {A.}~\bibnamefont {Alexander-Katz}},\
  }\href {\doibase https://doi.org/10.1016/j.bpj.2019.07.026} {\bibfield
  {journal} {\bibinfo  {journal} {Biophysical Journal}\ }\textbf {\bibinfo
  {volume} {117}},\ \bibinfo {pages} {892} (\bibinfo {year}
  {2019})}\BibitemShut {NoStop}%
\bibitem [{\citenamefont {Sciortino}\ \emph {et~al.}(2020)\citenamefont
  {Sciortino}, \citenamefont {Zhang}, \citenamefont {Gang},\ and\ \citenamefont
  {Kumar}}]{sciortino2020combinatorial}%
  \BibitemOpen
  \bibfield  {author} {\bibinfo {author} {\bibfnamefont {F.}~\bibnamefont
  {Sciortino}}, \bibinfo {author} {\bibfnamefont {Y.}~\bibnamefont {Zhang}},
  \bibinfo {author} {\bibfnamefont {O.}~\bibnamefont {Gang}}, \ and\ \bibinfo
  {author} {\bibfnamefont {S.~K.}\ \bibnamefont {Kumar}},\ }\href@noop {}
  {\bibfield  {journal} {\bibinfo  {journal} {ACS nano}\ }\textbf {\bibinfo
  {volume} {14}},\ \bibinfo {pages} {5628} (\bibinfo {year}
  {2020})}\BibitemShut {NoStop}%
\end{thebibliography}%

\end{document}

\preprint{APS/123-QED}

%\author{~~~~~~}
%\affiliation{Department of Physics, Sapienza Universit\`a di Roma, Piazzale Aldo Moro, 2, 00185 Rome, Italy}

%`` <-- use this symbol for left quotes
\author{Lorenzo Rovigatti} 
\affiliation{Department of Physics, {\textit Sapienza} Universit\`a di Roma, Piazzale A. Moro 2, IT-00185 Roma, Italy}
% \alsoaffiliation{CNR-ISC Uos Sapienza, Piazzale A. Moro 2, IT-00185 Roma, Italy}

\author{Francesco Sciortino}
\affiliation{Department of Physics, {\textit Sapienza} Universit\`a di Roma, Piazzale A. Moro 2, IT-00185 Roma, Italy}
%\SectionNumbersOn

%%%%%%%%%%%%%%%%%%%%%%%%%%%%%%%%%%%%%%%%%%%
%%%%%%%%%%%%%%%%%%%%%%%%%%%%%%%%%%%%%%%%%%%

\date{\today}

\title{Supplementary Information to ``Designing enhanced entropy binding in single-chain nano particles''}

\maketitle

\section{Models and Numerical Details}
\label{sec:details}

\subsection{Models}
In the main text we discuss three polymer  models.  Each polymer is composed by $N_m=254$ monomers  (24 reactive  and 230 inert monomers M).
Reactive  monomers are equally spaced by ten inert monomers each. The 24 reactive monomers can be of four types, which we label A, B and C and D. The first polymer ( (AAAA)$_6$ in the following) is composed only of A-type reactive monomers.  The second polymer ( (ABAB)$_6$ in the following) is composed by an alternate regular sequence of A-type and B-type reactive monomers.  Finally, the third polymer ( (ABCD)$_6$ in the following) is composed by an alternate regular sequence of A-B- C- and D-type reactive monomers.  The sequences are reported in Table~\ref{tab:sequences}  and pictorially represented in the snapshots of the main text. %Here we also discuss the effect of  the spacing between reactive monomers and  the polymer length.

\begin{table}[htp]
\begin{center}
\begin{tabular}{ |c|c| } 
 \hline
        Name & Sequence     \\
       \hline
       (AAAA)$_6$  &  ${(AM_{10}})_{23}A$  \\
       \hline
       (ABAB)$_6$ &  ${(AM_{10}BM_{10}})_{11}AM_{10}B$  \\
        \hline
        (ABCD)$_6$ &  $ (AM_{10}BM_{10}CM_{10}DM_{10})_5AM_{10}BM_{10}CM_{10}D$   \\
\hline
\end{tabular}
\caption{The three main polymers investigated in this work. \label{tab:sequences}
}
\end{center}
\end{table}

\subsection{Numerical Details: interaction potential}

%Each polymer is composed by a sequence of $N_m$ monomers.  
All monomers that are nearest-neighbours along the chain
(topologically bonded)   interact through the Kremer-Grest~\cite{kremer1990dynamics} 
force field, sum of a WCA potential and a FENE potential.
More precisely, defining $r$ as the distance between bonded monomers, 
\begin{equation}
V_{\rm WCA}(r)= 4 \epsilon   \left [ \left ( \frac{r}{\sigma} \right )^{12} - \left ( \frac{r}{\sigma} \right )^{6} \right ] +V_{\rm shift}~~~~ r<2^{1/6}
\label{eq:wac}
\end{equation}

\noindent
where $\epsilon$ and $\sigma$ are the units of energy and length respectively and $V_{\rm shift}$ is a constant  that brings $V_{\rm WCA}$ 
 to zero at $r=2^{1/6} \sigma$.  For $r>2^{1/6}~~~V_{\rm WCA}(r)=0$.
 The FENE potential is
\begin{equation}
V_{\rm FENE}(r)= - \frac{1}{2} K  d_0^2 \ln \left( 1-\frac{d_0}{r}\right)^2
%       tmp=1.0d0-(r/d0f)**2
 %      efene= -0.5d0*akf*d0f**2*dlog(tmp)
\end{equation}
where $d_0=1.5 \sigma$ and $K=30 \epsilon/\sigma^2$.
Non-bonded monomers interacts  via an excluded-volume interaction, modeled with the same WCA potential.
The only attractive contribution arises from the interaction between the reactive monomers.

To model binding,  we borrow a functional form proposed by
Stillinger and Weber~\cite{stillinger1985computer} in their model for Silicon. The corresponding interaction potential  $V_{\rm bind}(R)$  is
\begin{equation}
V_{\rm bind}(r) = A \epsilon_{b} \left [ B  \left  ( \frac{\sigma_{s}}{r} \right )^{4} -1 \right ] e^{\sigma_{s}/(r-r_c)}
\label{eq:sw}
\end{equation}

where $\sigma_s=1.05 \sigma$, $r_c=1.68 \sigma$, $B= 0.41$, $A=8.97$.  The coefficient $\epsilon_{b}$ modulates the
strength of the binding attraction.  $V_{\rm bind}$ goes continuously to zero as $r$ approaches $r_c$.  In what follows, the attractive potential acts only between reactive monomers of the same type.

To enforce the single-bond per reactive monomer condition we implement a repulsive three body interaction $V_{3b}$ which
penalizes the formation of triplets of bonded monomers~\cite{sciortino2017three}.  
 In addition, $V_{3b}$ is designed to almost exactly compensate the gain associated to the formation of a second bond,
 originating an almost flat energy hyper-surface which favours bond swapping even in the presence of 
 pair attraction energies much larger than the thermal energy. The   $V_{3b}$ interaction potential reads

\begin{equation}
V_{3b}(r_{ij},r_{ik})= 0.9 \epsilon_{b}  \sum_{ijk}      V_{3}(r_{ij}) V_{3}(r_{ik})
\label{eq:tb}
\end{equation}
where the sum runs over all triplets of bonded particles  (monomer $i$ bonded both with $k$ and $j$). 
$r_{ij}$ is the distance between particle $i$ and $j$.  The pair potential $V_{3}(r)$  is  defined in terms of the normalized $V_{\rm bind}(r)$ as

 \begin{equation}
 \label{eq:3body}
     V_3(r) =  \left\{ \begin{array}{lr}
        1  &  r\leq r_{min} \\
        -\frac{V_{\rm bind}(r)}{\epsilon_{b}},              &r_{min} \leq  r\leq r_{c}         
        \end{array}  \right \}  
 \end{equation} 
 
 where $r_{min}$ is the distance at which  $V_{\rm bind}(r)$ has a minimum.  We note on passing that no additional 
computational resources are requested to calculate $V_{3b}(r_{ij},r_{ik})$, since
the latter is defined in terms of previously calculated quantities.  We also note that differently from the Stillinger-Weber
potential~\cite{stillinger1985computer}, which favours the formation of a tetrahedral ordering  via an angular dependence,  here the three-body potential does not depend on $r_{jk}$.   

\begin{figure}[htbp] %  figure placement: here, top, bottom, or page
   \centering
      \includegraphics[width=0.45\textwidth]{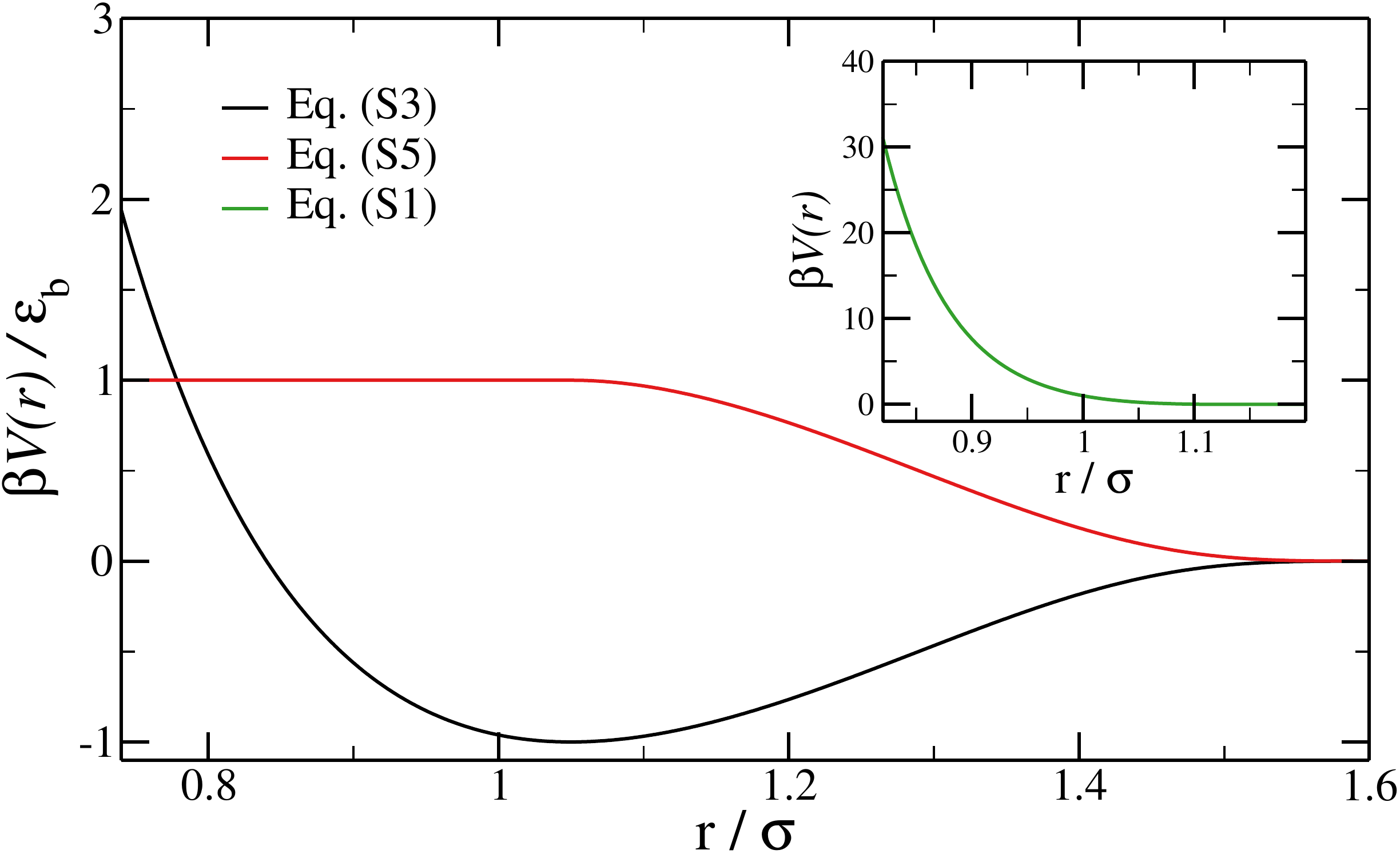} 
   \caption{The different potentials acting between non-bonded monomers.}
   \label{fig:pot}
\end{figure}

Fig.~\ref{fig:pot} shows  the shape of $V_{\rm WCA}$ (Eq.~\ref{eq:wac}), $V_{\rm bind}(r)$ (Eq.~\ref{eq:sw}), and $V_3(r)$ (Eq.\ref{eq:3body}).

\subsection{Numerical Details: Effective potential calculations}
\label{sec:effpotcalc}
We have calculated  the effective potential performing simulations of two identical polymers 
at $k_BT=1$ for $12<\epsilon_b<17$
constrained
by a harmonic potential acting on the relative distance  $R$ between their centres of mass $R  \equiv | \vec r^{CM}_1-\vec r^{CM}_2|$ and centred around $R_0$,
\begin{equation}
V_{\rm umbrella}(R)=\frac{1}{2}K (R-R_0)^2
\end{equation}
We have explored 15 different $R_0$ values, spaced from $R_0=\sigma$ to $R_0=30 \sigma$. For each simulation $i$ we have computed the probability distribution $P(R)$ that the two centres of mass are at a distance $R$.

% so that

From $P(R)$ the effective potential $\beta V_{\rm eff}(R)$ can be estimated as

\begin{equation}
\beta V_{\rm eff}(R) \sim -\ln \left( \frac{P(R)}{R^2}  e^{\beta V_{\rm umbrella}(R)}\right)
\end{equation}

\begin{figure}[!htbp] %  figure placement: here, top, bottom, or page
   \centering
   \includegraphics[width=3in]{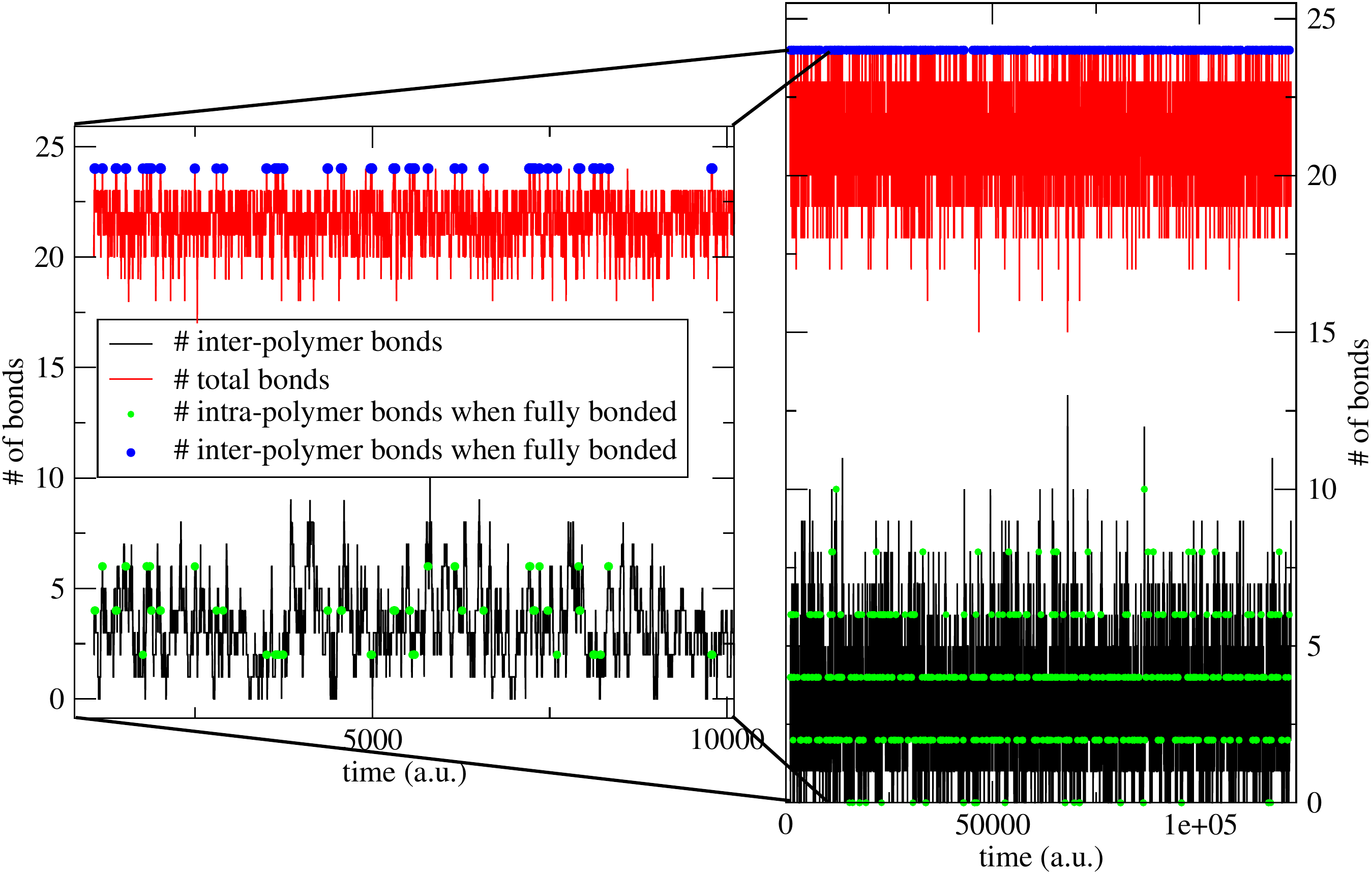} \\
   \includegraphics[width=3in]{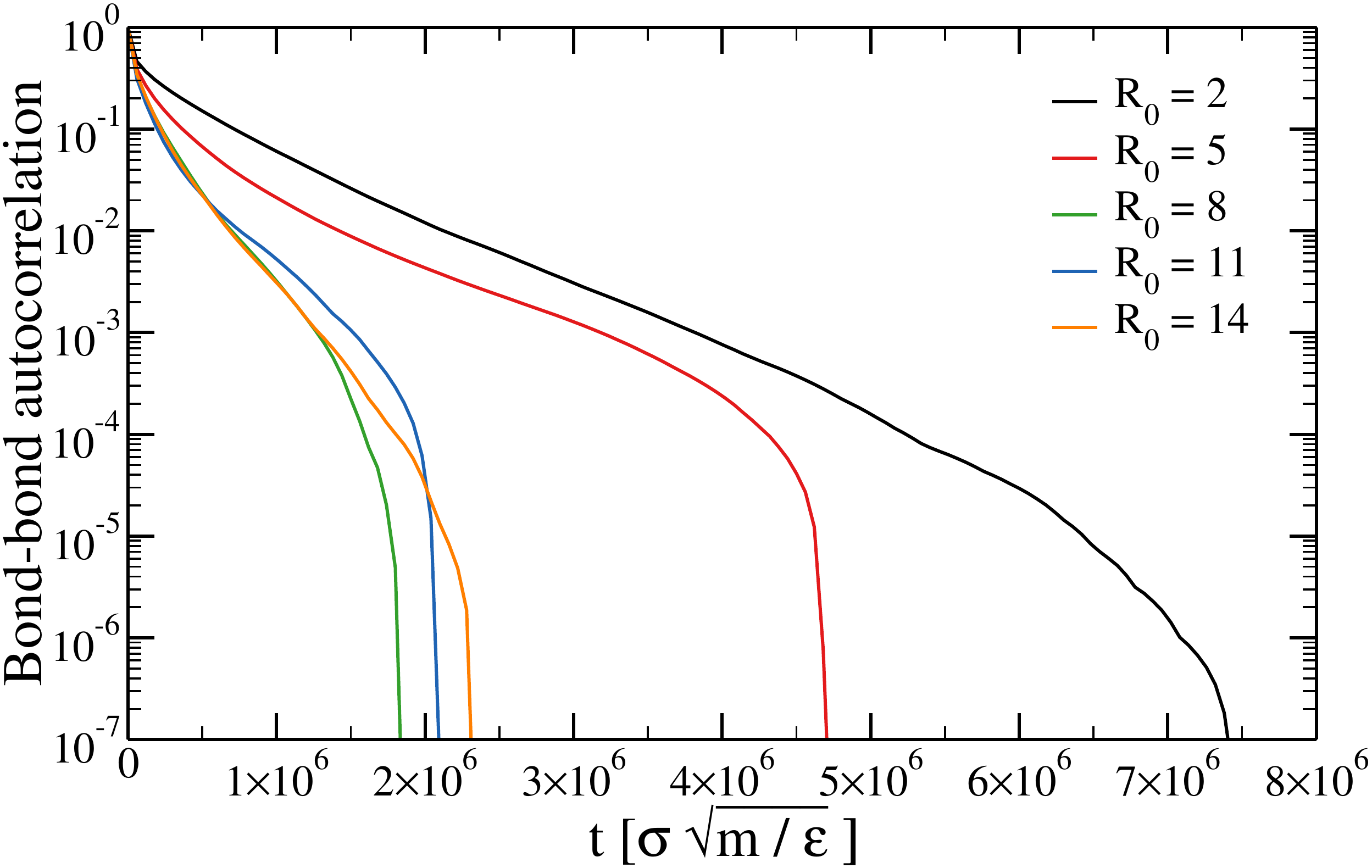} 
   \caption{Top: Time dependence of the total number of bonds (intra and inter) and of the intra bonds for two $AAAA$ polymers fixed at  center of mass distance 10 $\sigma$ when $\beta \epsilon_b=13$. The left panel is a zoom-in of the right panel that highlights the evolution of the time series. Bottom: The bond autocorrelation function for the ABCD system for selected umbrella sampling windows. Note that the overall length of the simulations is roughly four times longer than the time interval shown here ($\approx 3.2 \times 10^{7} \, \sigma \sqrt{m / \epsilon}$).}
   \label{fig:bondvstime}
\end{figure}

We extract from the simulation the configurations in which all  possible bonds are formed.  Only these configurations are   included in the statistical average. The top panel of Figure~\ref{fig:bondvstime} shows how the bonding pattern evolves during a simulation for the $R = 10 \, \sigma$ window of the $AAAA$ polymer: the number of bonds fluctuate and fully-bonded configurations are obtained repeatedly.  By extracting the fully-bonded configurations it is possible to follow the time evolution of the
   inter-polymers bonds (green dots), in fully bonded states. Similar results are obtained for all investigated  center of mass distances and for all polymer models. 
   
The bottom panel of Figure~\ref{fig:bondvstime} shows the bond autocorrelation function, defined as the probability that a bond that is present at $t = 0$ also exists, without ever being broken, at a later time $t$, as computed for selected umbrella sampling windows in the whole range of explored chain-chain separations for the ABCD system, which is the system with the slowest convergence rate of all the systems. For all explored windows, the probability that a bond exists uninterruptedly for $\approx 8 \times 10^{6} \, \sigma \sqrt{m / \epsilon}$ is always less than $10^{-7}$. We note that our simulations run for at least four times longer than the time range shown here, meaning that the system is able to completely forget about its initial bonding pattern several times during the course of the simulation.

For each of the three models 
we evaluated two effective interactions: the full potential, $V_{\rm eff}(R)$, and the potential between polymers that can bind only intramolecularly, $V_{\rm eff}^{\rm intra}(R)$.  In this latter case the calculated potential provides a measure of the repulsive excluded volume contribution.  The difference $V_{\rm eff}(R)-V_{\rm eff}^{\rm intra}(R)$ provides a measure of the entropic attraction between the two polymers. This entropic contribution is the sum of the combinatorial and conformational terms.

\noindent
Result from different simulations have been combined together with the WHAM method~\cite{ferrenberg1989optimized}.

\subsection{Numerical Details: Equation of state calculations}
\label{sec:eos}

For the (AAAA)$_6$, (ABAB)$_6$ and (ABCD)$_6$ models we have run bulk molecular dynamics simulations at constant volume $V$, temperature $T$ and number of polymers $N_c$. We use a velocity Verlet integrator with time step $\delta t = 0.003$ in units of $\sigma\sqrt{m / \epsilon}$, where $m$ is the mass of a monomer, and a Andersen-like thermostat~\cite{russo2009reversible}. We fix $k_B T = 1$, $N_c = 100$ and vary $V$ to compute the equation of state in the required density range.

\subsection{Numerical Details: Direct coexistence calculations}
\label{sec:directcoex}

We have also performed direct-coexistence molecular dynamics simulations in which we use an elongated box size along the $x$ axis containing $N_c = 200$ chains. We use the same simulation parameters of Section~\ref{sec:eos}.

Since each reactive monomer can be involved in a single bond only, the maximum number of bonds that can be formed is given by $N_b^{\rm max} = 24 N_c/2 $. We define $p_b = N_b / N_b^{\rm max}$, where $N_b$ is the number of formed bonds. In the simulations there are never more than 2 unformed bonds (\textit{i.e.} $p_b > 0.993$), so that we can safely assume that the energy is constant and the only driving force is entropic.

\subsection{Numerical Details: Hamiltonian integration}
\label{subsec:HI}

We evaluate the entropy difference  $ \Delta S_{\rm o \rightarrow fb}$ between the open and fully-bonded state for the (AAAA)$_6$ , (ABAB)$_6$ and (ABCD)$_6$ SCNP. To do so we run several simulations of single polymers at 21 equally spaced values of the binding strength $\epsilon_{b}$ ranging from $\epsilon_{b} = 0$, which corresponds to an open state (identical for all polymers), to $\epsilon_{b} = 20$, corresponding to a fully bonded state (different for all polymer types). The change in free energy between the two states is given by
\begin{equation}
\label{eq:hamiltonian_integration}
\Delta F_{\rm o \rightarrow fb} = \int_0^{20} \left \langle \frac{U_b}{\epsilon_{b}} \right \rangle d\epsilon_{b}
\end{equation}
where $U_b$ is the total binding energy of a configuration and $\left \langle \cdot \right \rangle$ is an ensemble average. Since, by definition, $\Delta F_{\rm o \rightarrow fb} = \Delta U_{\rm o \rightarrow fb} - T \Delta S_{\rm o \rightarrow fb}$, where $\Delta U_{\rm o \rightarrow fb}$ is the difference in the average total energy between the two states, the entropy difference between the open and closed state of a given polymer is
\begin{equation}
\label{eq:hamiltonian_entropy}
\Delta S_{\rm o \rightarrow fb} = \frac{\Delta U_{\rm o \rightarrow fb} - \Delta F_{\rm o \rightarrow fb}}{T}.
\end{equation}

\begin{figure}[htbp] %  figure placement: here, top, bottom, or page
   \centering
         \includegraphics[width=0.45\textwidth]{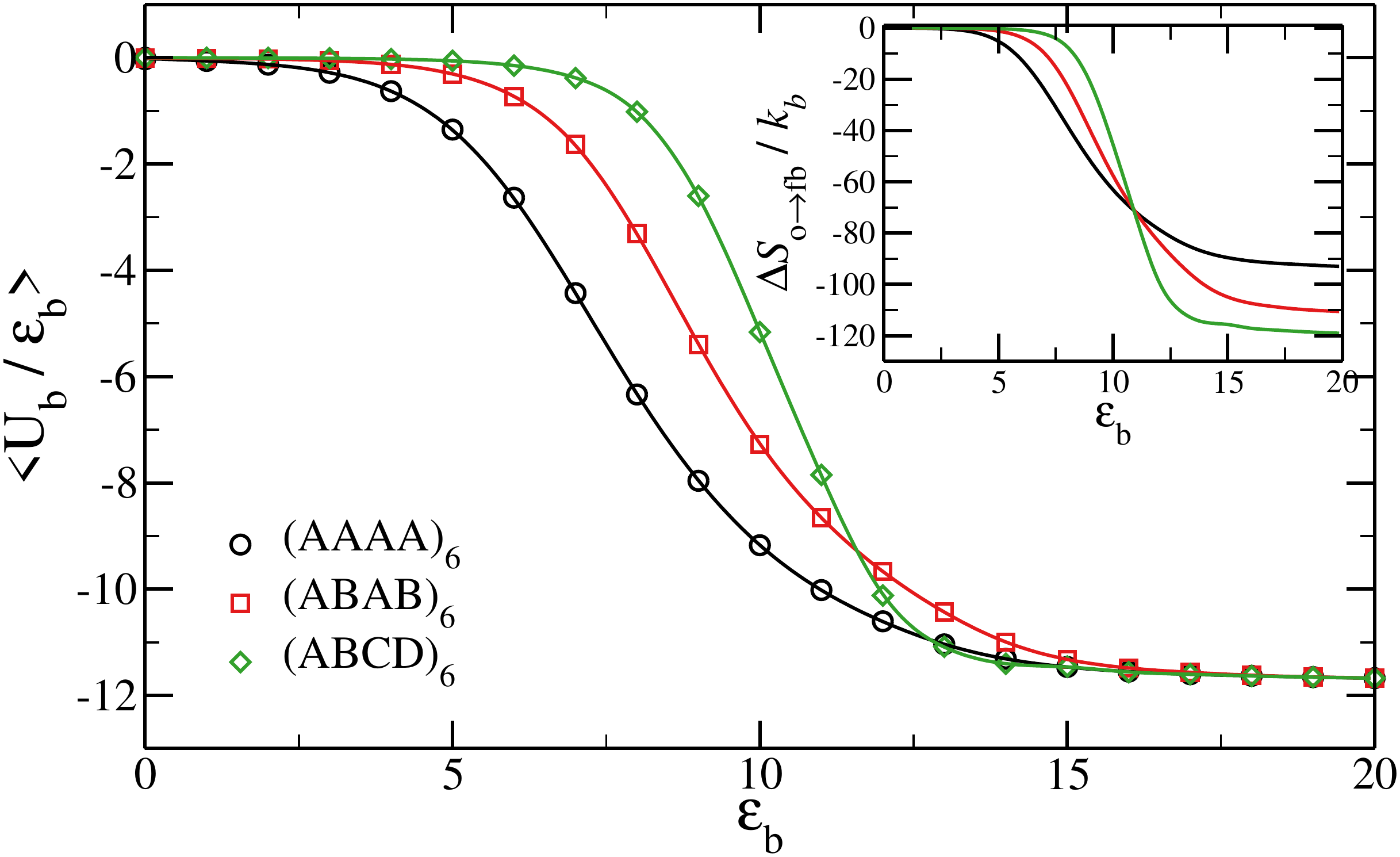} 
   \caption{The integrand of Eq.~\ref{eq:hamiltonian_integration} as a function of the two-body interaction strength $\epsilon_b$ for the three polymer models studied in the main text. The inset shows the resulting entropy difference computed through Eq.~\ref{eq:hamiltonian_entropy}.}
   \label{fig:hamiltonian_integration}
\end{figure}

Figure~\ref{fig:hamiltonian_integration} shows $\left \langle \frac{U_b}{\epsilon_{b}} \right \rangle$ and $\Delta S_{\rm o \rightarrow fb}$ for the three polymer models investigated in the main text.

\section{Additional polymers}
\label{sec:varying}

\subsection{Varying the spacing between the reactive sites}
\label{sec:varyingX}

\begin{table*}[!htbp]
\begin{center}
\begin{tabular}{ |c|c|c| } 
 \hline
        Name & Sequence & $R_g^2 / \sigma^2$     \\
       \hline
       (AAAA)$_6$ (M=20)  &  ${(AM_{20}})_{23}A$  &  $161$  \\
       \hline
       (ABAB)$_6$ (M=20) &  ${(AM_{20}BM_{20}})_{11}AM_{20}B$  &  $110$  \\
        \hline
        (ABCD)$_6$ (M=20) &  $ (AM_{20}BM_{20}CM_{20}DM_{20})_5AM_{20}BM_{20}CM_{20}D$   &  $108$ \\
               \hline
       (AAAA)$_3$ (M=20)  &  ${(AM_{20}})_{11}A$  &  $85$  \\
       \hline
       (ABAB)$_3$ (M=20) &  ${(AM_{20}BM_{20}})_{5}AM_{20}B$  &  $60$  \\

\hline
\end{tabular}
\caption{The three  polymers  with larger spacing between the reactive sites. \label{tab:sequencesM20}
}
\end{center}
\end{table*}

We evaluate here the effect of the spacing between reactive sites by comparing the same three polymers studied in the article
with three additional polymers with the very same number of reactive sites (24), but spaced by $M=20$ inert monomers (instead of $M=10$).  The total length of these additional three polymers is thus  $N_m=484$ monomers  (24 reactive  and 460 inert monomers M).
The gyration radius of the three new polymers is reported in Table~\ref{tab:sequencesM20}.

\begin{figure}[htbp] %  figure placement: here, top, bottom, or page
   \centering
         \includegraphics[width=0.45\textwidth]{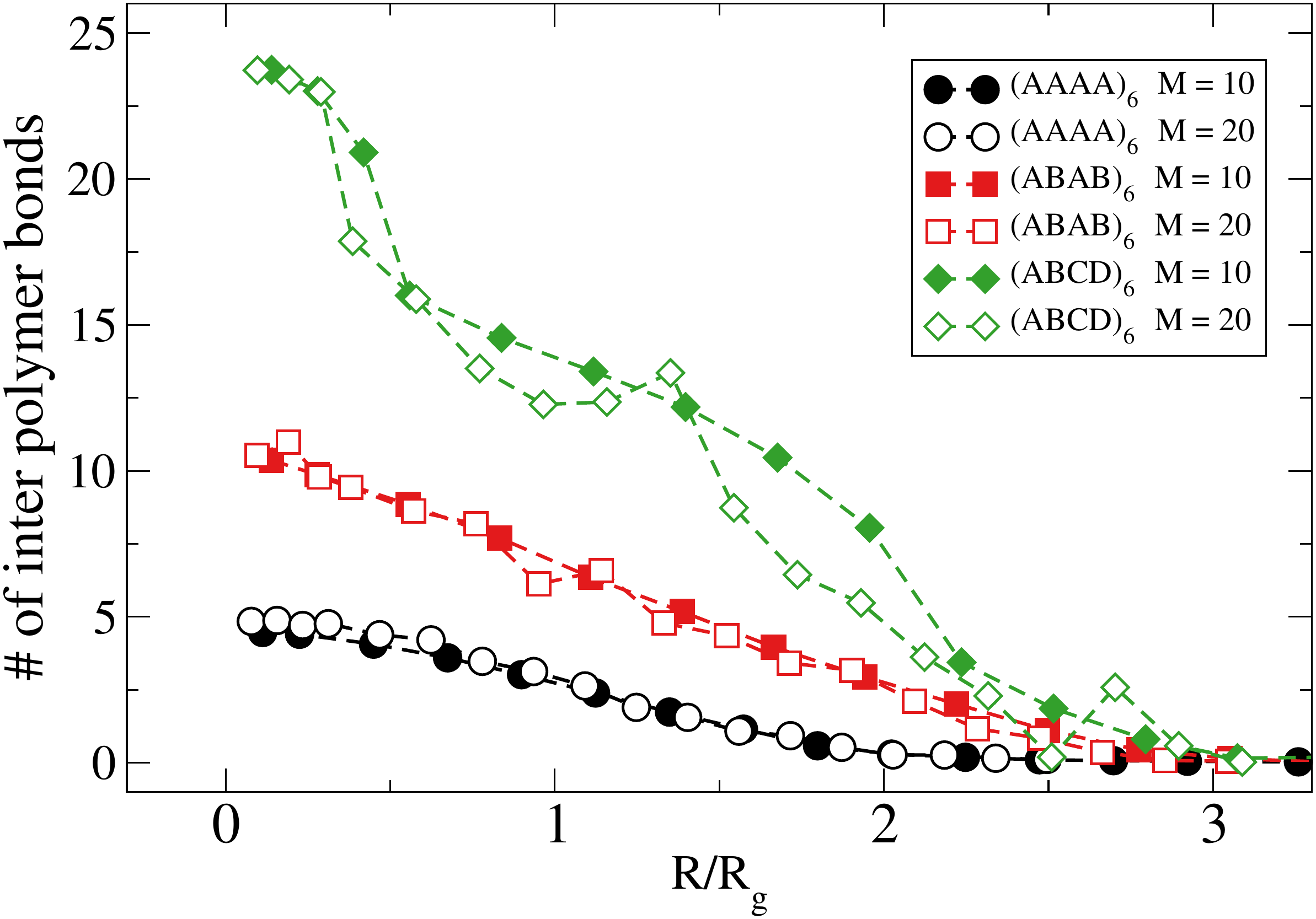} 

      \includegraphics[width=0.45\textwidth]{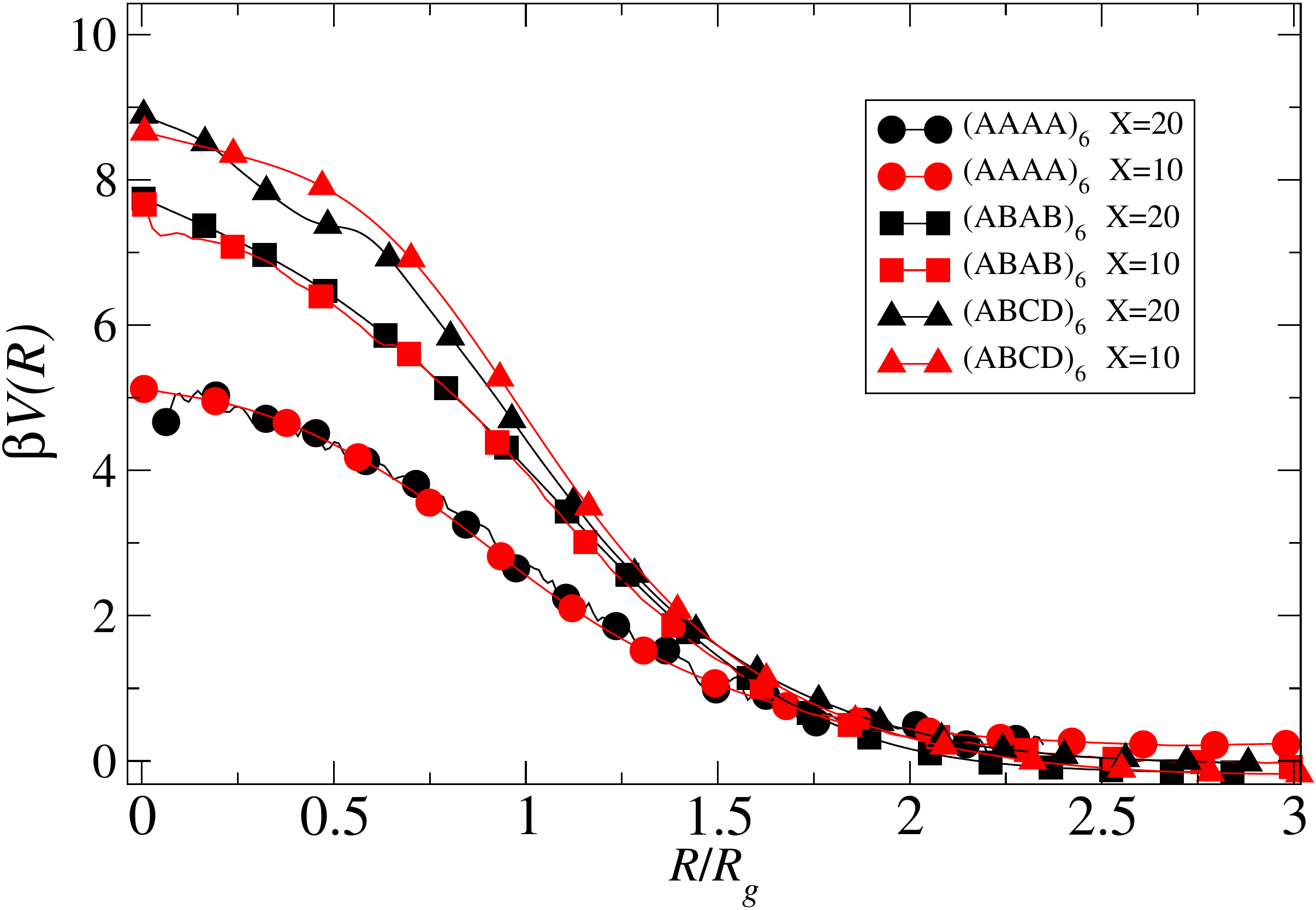} 

      \includegraphics[width=0.45\textwidth]{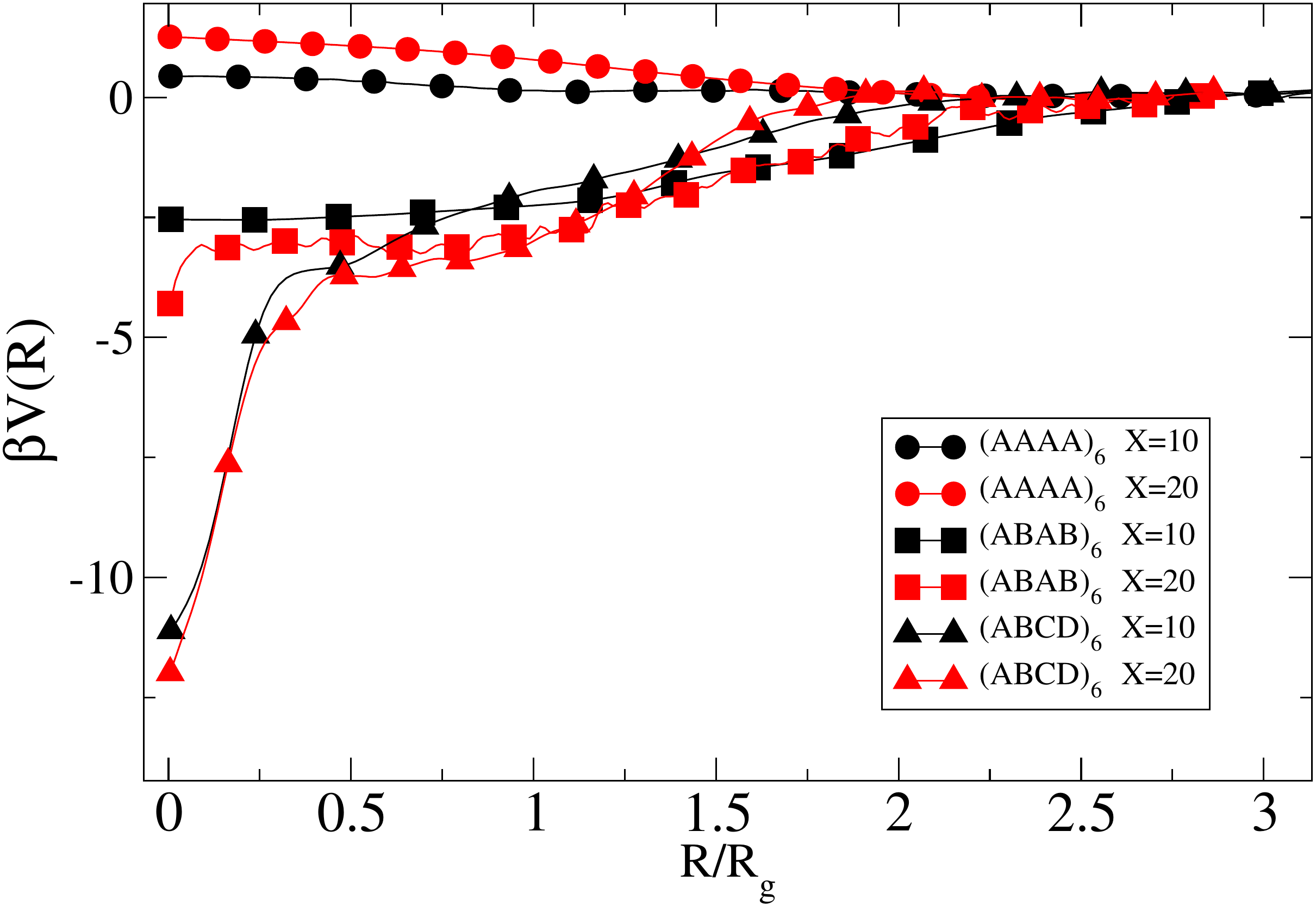} 
   \caption{Comparison of the number of inter-polymer bonds, of the only-intra bonds effective potential 
   $\beta V_{eff}^{intra}$, 
   and of the effective potential $\beta V_{eff}$ vs the relative distance $R$ between the two centers of mass, rescaled by  the respective gyration radii.}
   \label{fig:spacingM}
\end{figure}

Figure~\ref{fig:spacingM} shows that, when plotted as a function of $R/R_g$, the number of inter-polymer bonds, the only-intra potential and the effective potential, are seemingly independent (or weakly dependent) of $M$.

\begin{figure}[!htbp] %  figure placement: here, top, bottom, or page
   \centering
   \includegraphics[width=0.5\textwidth]{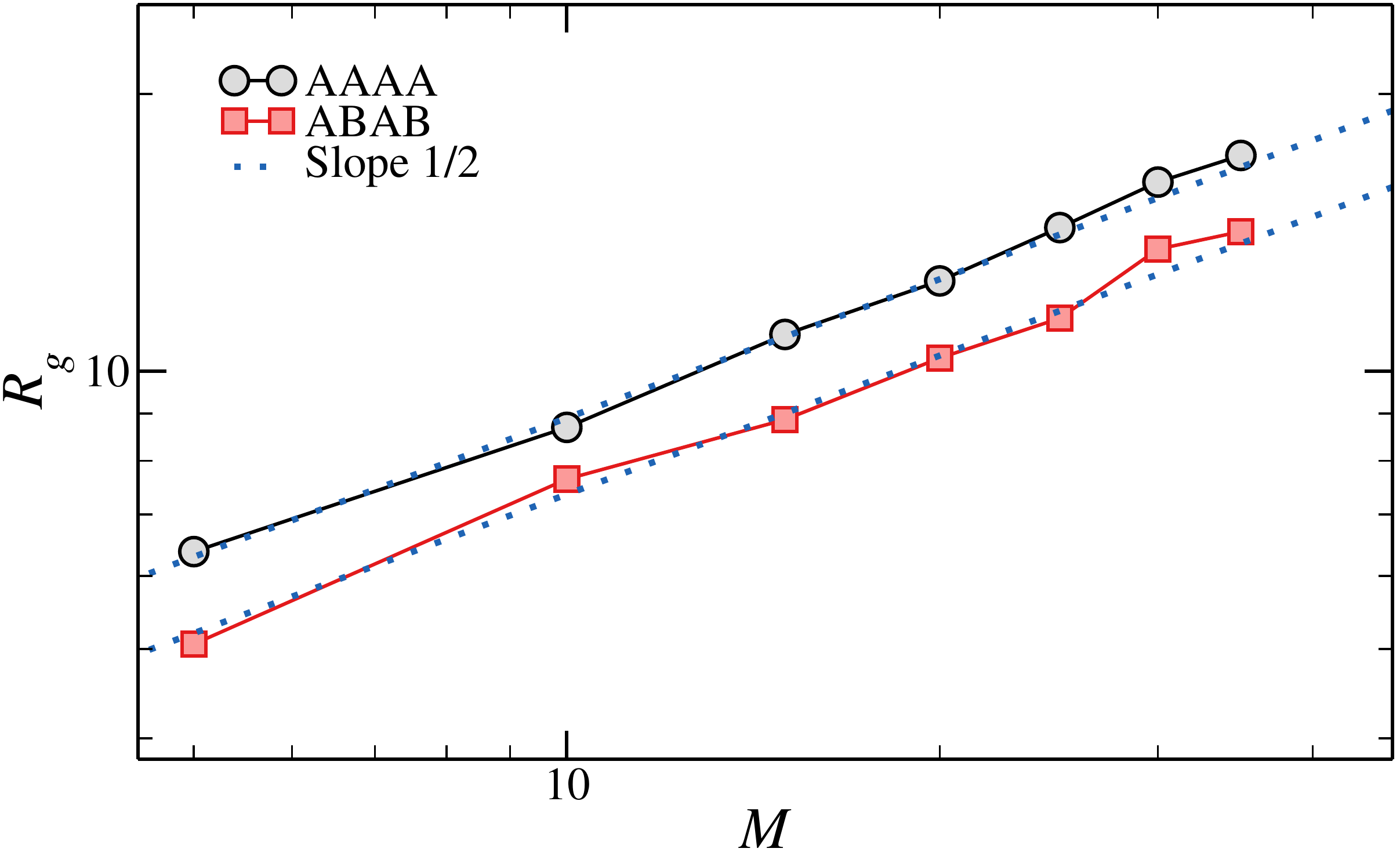} 
   \caption{The gyration radius of single AAAA and ABAB polymers as a function of $M$.}   
   \label{fig:Rg}
\end{figure}

Data on irreversible single-chain nanoparticles show that the scaling exponents of irreversible fully-flexible SCNPs are those of Gaussian chains~\cite{moreno2017effect,formanek2017effects}. Figure~\ref{fig:Rg} shows that this is the case also for fully-bonded, reversible SCNPs, as the dependence of the radius of gyration of both AAAA and ABAB polymers on $M$ is compatible with an exponent $0.5$. Data of Ref.~\cite{moreno2017effect} also shows that the gyration radius of the loops (which is related to the entropic cost of forming a loop) scales with $M$ with apparent exponents that depend on the loop size. As a result, understanding the reason for the weak $M$ dependence is not straightforward and simple scaling (mean-field-like) arguments like those of Ref.~\cite{semenov1998thermoreversible} need a finer tuning.

Regardless of its origin, the consequence on the thermodynamics of this scaling can be stated by saying that the 
phase diagram, expressed as a function of the temperature and of the scaled density $\rho R_g$, should be independent on the length of the polymer at fixed number of reactive sites.  The number of spacing monomers is thus irrelevant. 
This consideration also indicates that the $(AAAA)_6$ polymers will never phase separate, independently from the
number of spacing monomers, as predicted by Semenov and Rubinstein~\cite{semenov1998thermoreversible}.

\subsection{Varying the degree of polymerization}
\label{sec:varyingNc}

\begin{figure}[!h] %  figure placement: here, top, bottom, or page
   \centering
         \includegraphics[width=0.45\textwidth]{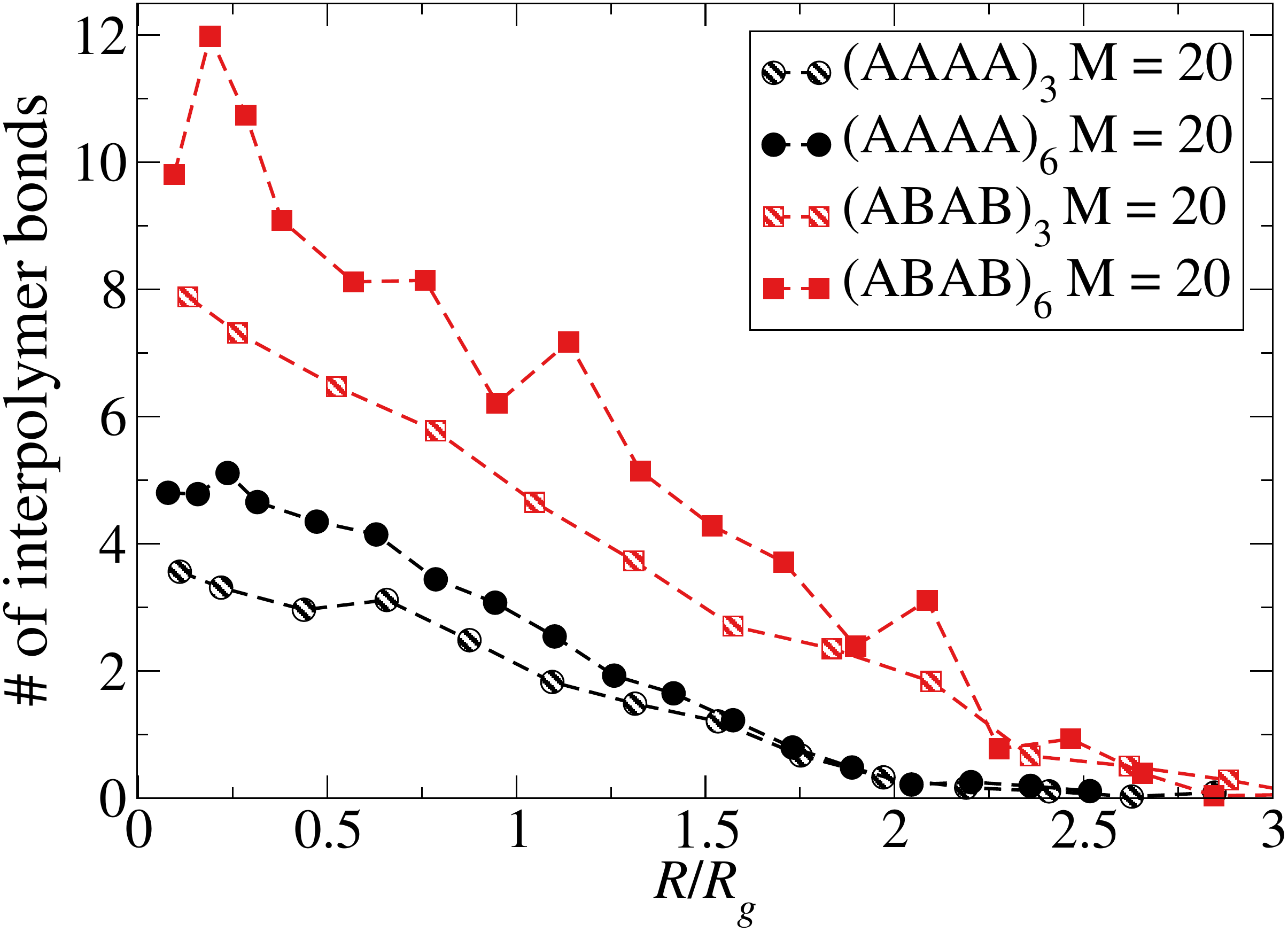} 
%         \vskip -1.cm
 %     \includegraphics[width=0.45\textwidth]{scaled-vrAA-2.pdf} 
  %             \vskip -1.cm
  %    \includegraphics[width=0.45\textwidth]{scaled-vrAB-2.pdf} 
   \caption{Comparison of the number of inter-polymer bonds
   %, the only-intra bonds effective potential and the effective potential 
   vs the relative distance between the two center of mass, scaled by  the respective gyration radius.}
   \label{fig:spacingR}
\end{figure}

We evaluate here the effect of the degree of  polymerization at fixed number of  spacing inert monomers  ($M=20$)
on changing the number of repeating units (and hence of the reactive monomers). Specifically, polymers
are constituted by 12 reactive units, for the AAAA and ABAB cases.
The gyration radius of the two new polymers is also reported in Table~\ref{tab:sequencesM20}. 
Fig.~\ref{fig:spacingR} compares the number of inter polymer bonds for 
polymers with 12 and 24 reactive sites with the same number of inert monomers.  Differently from the previous case, 
when distances are scaled by $R_g$, the number of reactive site within $R_g$ is halved.  Hence, the shorter polymer has  a lower probability to
form inter-particle bonds, as reflected in the figure.

\section{Conformational and combinatorial entropies}

\begin{figure}[!htbp] %  figure placement: here, top, bottom, or page
   \centering
   \includegraphics[width=0.48\textwidth]{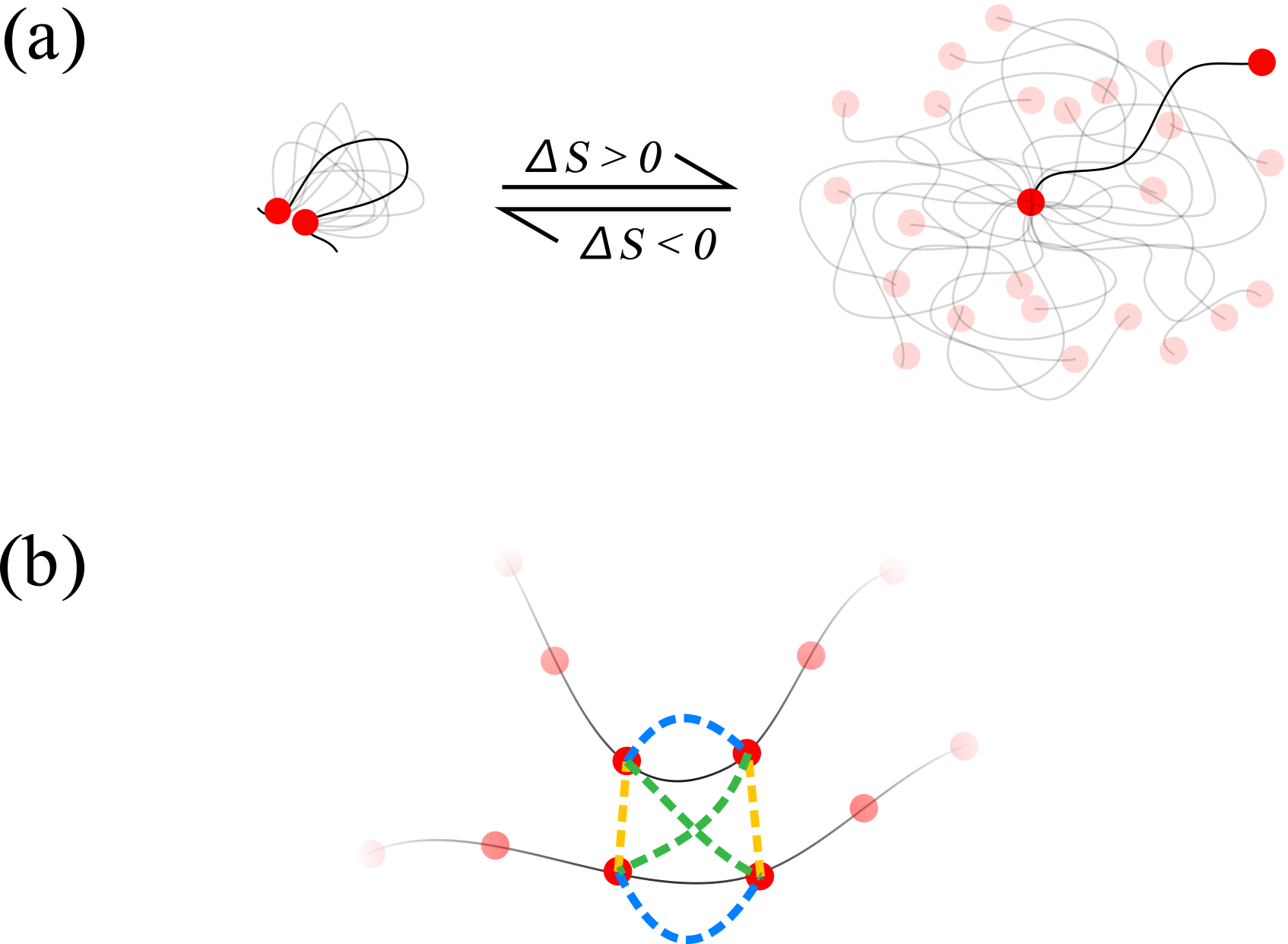} 
   \caption{Sketches providing visual representations of some of the different entropic contributions that play a role in the thermodynamics of fully-bonded SCNPs.  
   Chains are drawn as black lines decorated with red spheres,  representing the reactive monomers.  The grey lines indicate possible different local configurations, one of which is highlighted in black. 
    (a) Closing a loop costs conformational entropy, since the additional constraint greatly limits the available chain configurations (black and shaded curves). (b) In the fully-bonded limit, when sites belonging to different chains come in contact there are multiple ways of satisfying all the bonds (here shown with dotted lines). If we focus on the four attractive sites in the middle of the picture we see that there is only one possible way of forming intra-molecular bonds (blue lines), but two ways of forming inter-molecular bonds (green and yellow lines), which are thus favoured from a purely combinatorial point of view.}
   \label{fig:entropies}
\end{figure}

Figure~\ref{fig:entropies} provides visual representations of the entropic contributions due to conformational and combinatorial entropies. Note that for the latter in general the different bonding patterns (see Figure~\ref{fig:entropies}(b)) also have different conformational contributions, since they are composed by different numbers of inter and intra-molecular bonds. In the next section we explicitly compute the combinatorial entropy of a system composed of two polymers decorated with reactive monomers that can form both intra- and inter-molecular bonds, with the crucial assumption that all bonds have the same statistical weight. For chain polymers this is a strong approximation (see \textit{e.g.} Ref.~\cite{ZUMBRO2019892}), but the derivation below can be nonetheless a useful starting point for a more rigorous approach.

\section{Evaluating the combinatorial entropy in a simple system}
\label{sec:comb_entropy}

Assume that  two polymers are at a  distance compatible for inter-polymer binding and that at this distance
there are $2N$ reactive monomers available to form $N$ bonds. Assume
that the first $N$ reactive monomers  belong to the first polymer and that the second half belong to the second polymer.
The total number of different bonding patterns is $(2N-1)!!$ (where $!!$ indicates the double factorial), since
the first picked reactive monomer can bind with $2N-1$ other, the second picked  can bind with $2N-3$  and so on.
We note that the number of states with only intra-bonds (zero inter bonds) is  simply given by $[(N-1)!!]^2$, since the first reactive monomer can bind only to $N-1$ others and so on, generating a
$(N-1)!!$ contribution for each of the two polymers.

Thus two polymers allowed to form inter-polymer bonds are stabilized, compared to two isolated polymers in which only intra-bonds are possibile by an entropic factor~\cite{sciortino2020combinatorial} 

\begin{equation}
\frac{\Delta S_{\rm comb}}{k_B} =
 \ln \left [ \frac{(2N-1)!!}{ [(N-1)!! ]^2}
 \right ]
 \label{eq:scomb}
\end{equation}

\begin{figure}[htbp] %  figure placement: here, top, bottom, or page
   \centering
      \includegraphics[width=0.45\textwidth]{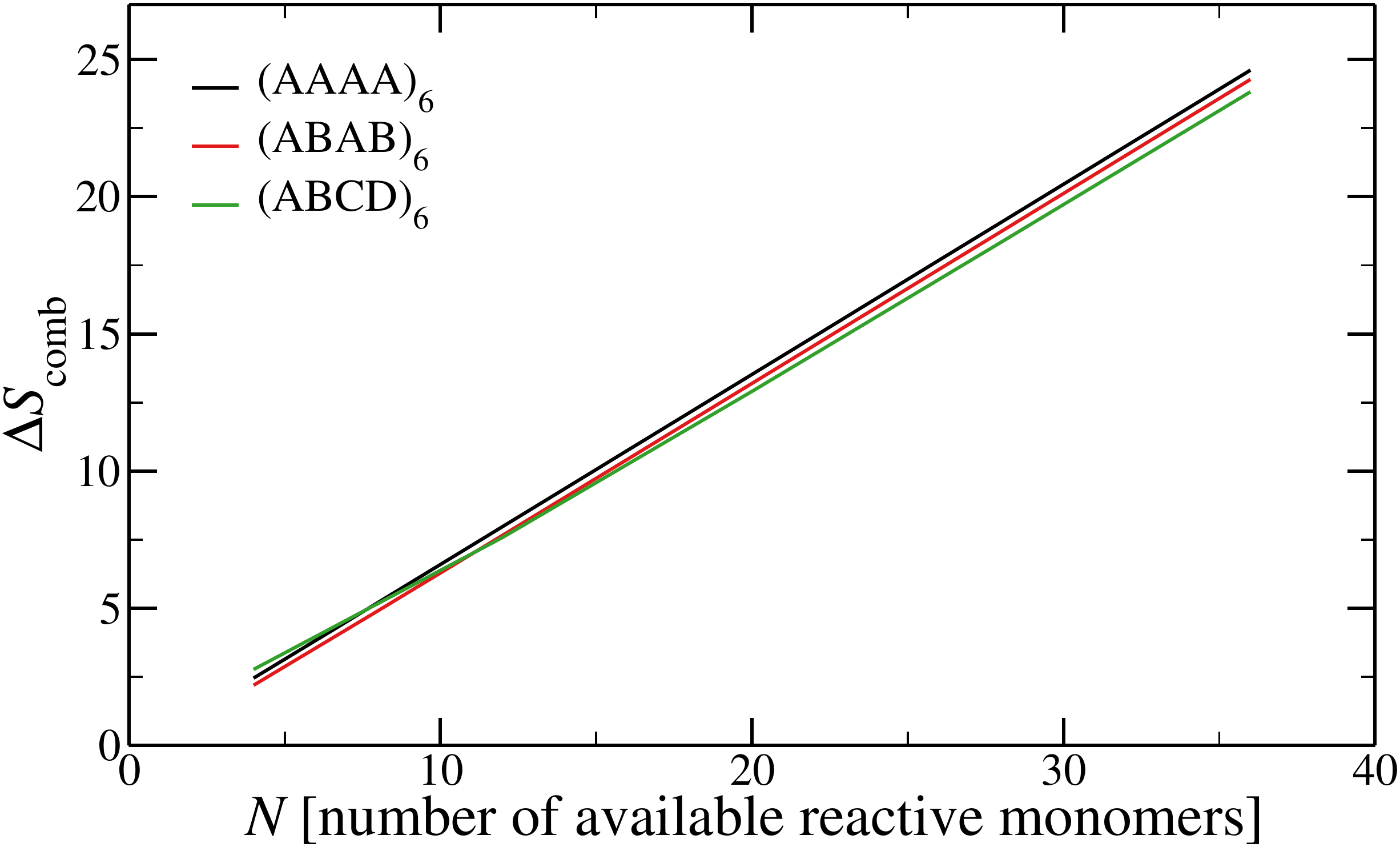} 
   \caption{Combinatorial entropic gain associated to inter-polymer binding for the three polymer configurations.  All polymers are characterized by comparable entropies. }
   \label{fig:dsteo}
\end{figure}

If reactive monomers of different type are present, as in the ABAB and in the ABCD SCNP,  the entropic difference can be
generalized as
\begin{equation}
\frac{\Delta S_{\rm comb}}{k_B} = \sum_{\alpha=1}^{m_t}
 \ln \left [ \frac{(2N_\alpha-1)!!}{ [N_\alpha-1)!! ]^2}
 \right ]
 \label{eq:scombalpha}
\end{equation}
where now $N_\alpha$ indicates the number of reactive sites of type $\alpha$, $m_t$ is the number of types and $N=\sum_\alpha N_\alpha$. 
Fig.~\ref{fig:dsteo} shows the resulting contribution of the combinatorial entropy  $S_{\rm comb}$ vs $N$ for the three different polymer topologies.   $\Delta S_{\rm comb}$ shows somewhat surprisingly a very weak dependence on  topology, with a slight preference for $AAAA$ over $ABAB$ and $ABCD$ within a fraction of $k_B$. If $\Delta S_{\rm comb}$ were the only contribution, bonding would be just a little bit stronger for the $AAAA$  polymer.

The limit for large $N$ is particularly simple. 
It can be demonstrated  by noting that the expression in Eq.\eqref{eq:scomb} can be recast by using the relations
$$
(2N - 1)!! = \frac{(2N)!}{2^N N!}
$$
and
$$
(N - 1)!! = \frac{N!}{2^{N/2}(N/2)!}
$$
so that 
$$
\frac{(2N - 1)!!}{[(N - 1)!!]^2} = \frac{(2N)!}{N!N!} \frac{(N/2)!(N/2)!}{N!} = \frac{\binom{2N}{N}}{\binom{N}{N/2}}.
$$
We now note that
$$
\binom{2N}{N} = \frac{2(2N - 1)}{N} \binom{2(N - 1)}{N - 1}.
$$
By applying this relation $N/2$ times we find
$$
\binom{2N}{N} = 2^{N/2} \prod_{i=0}^{N/2 - 1} \frac{2(N - i) - 1}{N - i} \binom{N}{N / 2}
$$
and therefore
$$
\frac{(2N(R) - 1)!!}{[(N(R) - 1)!!]^2} = 2^{N/2} \prod_{i=0}^{N/2 - 1} \frac{2(N - i) - 1}{N - i} 
$$
$$
= 2^{N/2} \prod_{i=0}^{N/2 - 1} 2 \left( 1 - \frac{1}{2(N - i)} \right) = 2^{N} \prod_{i=0}^{N/2 - 1} \left( 1 - \frac{1}{2(N - i)} \right).
$$
The combinatorial entropy (Eq.~\ref{eq:scomb}) thus can be equivalently rewritten as 
$$
\frac{\Delta S_{\rm comb}}{k_B} = N\log 2 + \sum_{i=0}^{N/2 - 1} \log \left( 1 - \frac{1}{2(N - i)} \right).
$$
We can derive a simpler (approximated) expression if $N$ is large. Indeed, in this limit $\log \left( 1 - \frac{1}{2(N - i)} \right) \approx -\frac{1}{2(N - i)}$ and therefore
$$
\sum_{i=0}^{N/2 - 1} \log \left( 1 - \frac{1}{2(N - i)} \right) \approx -\frac{1}{2} \sum_{i=0}^{N/2 - 1} \frac{1}{N - i} 
$$
$$
\approx -\frac{1}{2} 
\int_{N/2}^{N} \frac{dx}{x} = -\frac{1}{2}\log 2
$$
so that we find
\begin{equation}
\label{eq:scomb_approx}
\frac{\Delta S_{\rm comb}}{k_B} \approx (N-0.5)\log 2,
\end{equation}
simply indicating  that each reactive monomer, if part of an  intra bond, select half of the possible bonding possibilities.  

In the general case of $m_t \ne 1$ the latter expression becomes
$$
\frac{\Delta S_{\rm comb}}{k_B} \approx (N - 0.5 m_t)\log 2 \approx N \log 2.
$$

\noindent
In other words, in the large-$N$ limit the combinatorial entropic attraction $\Delta S_{\rm comb}/{k_B} $ 
does not depend on the  types of reactive monomers decorating  the polymer chain but only on their total number.

\bibliography{SI.bib}